\documentclass[twoside,11pt]{article}
\usepackage[utf8]{inputenc} % allow utf-8 input
\usepackage[T1]{fontenc}    % use 8-bit T1 fonts
\usepackage{hyperref}       % hyperlinks
\usepackage{url}            % simple URL typesetting
\usepackage{booktabs}       % professional-quality tables
\usepackage{amsfonts}       % blackboard math symbols
\usepackage{nicefrac}       % compact symbols for 1/2, etc.
\usepackage{microtype}      % microtypography
\usepackage{lipsum}
\usepackage{fancyhdr}       % header
\usepackage{graphicx}       % graphics
\graphicspath{{media/}}     % organize your images and other figures under media/ folder
\usepackage{natbib}
\usepackage{amsmath}
\usepackage{amssymb}
\usepackage{comment}
\usepackage{listings}
\usepackage{amsthm}
\usepackage{siunitx}
\usepackage{algorithm}
\usepackage{algpseudocode}
\usepackage{subcaption}
\graphicspath{ {./images/} }
\usepackage{array}
\usepackage{paralist}
\usepackage{verbatim}
\usepackage{multirow}
\usepackage{rotating}
\usepackage{appendix}
\usepackage{sectsty}
\allsectionsfont{\sffamily\mdseries\upshape}
\usepackage{algorithmicx}
\usepackage{fancyvrb}
\title{Covariance Supervised Principal Component Analysis
}

\author{
  Theodosios Papazoglou \\
  \texttt{theo.papazoglou@connect.hku.hk} \\
  Department of Statistics \& Actuarial Sciences \\
  The University of Hong Kong \\
  Pokfulam Road, Hong Kong
  \and
  Guosheng Yin \\
  \texttt{gyin@hku.hk} \\
  Department of Statistics \& Actuarial Sciences \\
  The University of Hong Kong \\
  Pokfulam Road, Hong Kong
}

\usepackage{PRIMEarxiv}
\begin{document}
\maketitle
\begin{abstract}
Principal component analysis (PCA) is a widely used unsupervised dimensionality reduction technique in machine learning, applied across various fields such as bioinformatics, computer vision and finance. However, when the response variables are available, PCA does not guarantee that the derived principal components are informative to the response variables. Supervised PCA (SPCA) methods address this limitation by incorporating response variables into the learning process, typically through an objective function similar to PCA. Existing SPCA methods do not adequately address the challenge of deriving projections that are both interpretable and informative with respect to the response variable. The only existing approach attempting to overcome this, relies on a mathematically complicated manifold optimization scheme, sensitive to hyperparameter tuning. We propose covariance-supervised principal component analysis (CSPCA), a novel SPCA method that projects data into a lower-dimensional space by balancing (1) covariance between projections and responses and (2) explained variance, controlled via a regularization parameter.
The projection matrix is derived through a closed-form solution in the form of a simple eigenvalue decomposition. To enhance computational efficiency for high-dimensional datasets, we extend CSPCA using the standard Nyström method. Simulations and real-world applications demonstrate that CSPCA achieves strong performance across numerous performance metrics.
\end{abstract}

% keywords can be removed

\section{Introduction}
Originally proposed by \citet{Pearson1901LIII.-On-lines-},
principal component analysis (PCA) has become one of the most popular statistical methods to perform dimensionality reduction (DR). PCA seeks for an orthogonal projection of the data onto a lower-dimensional linear space such that the variance of the projected data is maximised. It is usually performed as a pre-processing step before applying other machine learning models for classification or regression and it is commonly used for feature extraction or visualisation. PCA is an unsupervised learning method as it does not make use of the response variables during learning. However, when the response variables are available, the principal components derived by PCA are not guaranteed to be informative of the response variable which can result in low prediction accuracy when the goal is prediction. Recently, supervised PCA (SPCA) methods have been developed to address this issue by incorporating the response variables in the learning procedure, usually in the form of an objective function similar to the one of PCA. SPCA methods have received great attention lately in bioinformatics where high-dimensional microarray datasets are available and researchers are interested in prediction of different phenotype values \citep{Chen2008Supervised-prin,Shigemizu2019Risk-prediction}. However, the application of SPCA methods is not limited there as it has also gained attention in spectral analysis \citep{Bin2013Supervised-prin} and image classification \citep{Xia2014Semi--Supervise}.

In this paper, we focus mostly on data extracted from DNA microarray experiments where the predictors are gene expression measurements. In this type of dataset, the number of predictors is higher than the number of observations, usually ranging from thousands to tens of thousands of genes, making the use of dimensionality reduction methods necessary as a pre-processing step prior to prediction models. The response variables can be either continuous (e.g. C-reactive protein levels) or binary (e.g. tumor/cancer type). 
%In our analysis, we use partial least squares (PLS) regression \citep{Wold2001PLS-regression:} as a baseline comparison for regression tasks and linear discriminant analysis (LDA) \citep{FISHER1936THE-USE-OF-MULT} as a baseline comparison for classification tasks. The core idea of the former, that is maximising the covariance between the response variables and the data, closely aligns with our proposal.

As far as we are concerned, the first to address the issue of SPCA were \citet{Bair2004Semi-Supervised} and \citet{Bairetal} to predict patient survival from gene expression data. Instead of applying PCA directly on all genes before applying a linear model, they suggest to first identify a subset of genes which is most correlated with the response variable through univariate regression models and then apply PCA on this subset of features. This method has inspired other SPCA methods such as iterative supervised principal component analysis (ISPCA) \citep{Piironen2018Iterative-Super}, where an iterative procedure of Bair's method is performed to derive the supervised principal components. \citet{Bairetal} also proposed an underlying latent variable model to support their SPCA proposal. Similar latent variable models have been proposed in the literature for both PCA and SPCA. \citet{Tipping2002Probabilistic-P} proposed a probabilistic version of PCA in the form of a generative latent variable model, where the maximum likelihood solution of the model coincided with the solution obtained through traditional PCA. \citet{Bishop1998Bayesian-PCA} further extended this model in the Bayesian setting which is commonly referred to as Bayesian PCA. Probabilistic PCA has also been extended for SPCA. Specifically, \citet{Yu2006Supervised-prob} proposed a latent variable model for supervised and semi-supervised learning settings, where different versions of EM algorithms are applied to extract the maximum likelihood solutions.

SPCA was first formulated as an optimization problem of an objective function by \citet{Barshan2011Supervised-prin}, who attempted to solve the problem of SPCA by making use of the Hilbert-Schmidt independence criterion (HSIC). In a universal reproducing kernel Hilbert space (RKHS), two variables are independent if and only if their HSIC is equal to 0. By using an empirical measure of HSIC, they proposed an optimization procedure of an objective function that involves the trace of a matrix. The main advantage of their proposal is that a closed-form solution, in the form of matrix eigenvalue decomposition, exists. They also proposed a kernelised version of their method for non-linear mappings of the data which can be solved using generalised eigenvalue decomposition. While their method enhances predictive accuracy, it falls short of preserving the variation explained by traditional PCA.

A more recent approach proposed by \citet{Ritchie2019Supervised-Prin} referred to as least squares supervised principal component analysis (LSPCA), addresses this issue by introducing a regularisation parameter in the objective function that seeks to trade-off between prediction error and variance explained by the projected data. This idea is extended in a probabilistic framework by connecting LSPCA to generalised linear models \citep{Ritchie2020Supervised-PCA:}. In contrast to \citet{Barshan2011Supervised-prin}, LSPCA's optimization problem has no closed-form solution and its performance is sensitive to the choice of the regularisation parameter. In order to derive the projection from LSPCA, one has to perform gradient-based algorithms over manifolds.

We also discuss two common methods applied for dimensionality reduction when the response variables are available which are not part of the SPCA methods but can serve as baselines for the greater scope of supervised dimensionality reduction. The first is partial least squares (PLS) regression \citep{Wold2001PLS-regression:} which is mostly used for regression tasks and its objective aligns with our proposal. PLS is usually viewed as an extension to PCA as it aims to find a linear subspace of the data which captures the maximum covariance between the response variables and the data. In contrast to PCA that only projects the data into a lower dimensional space, PLS also projects the response variables in a new subspace. The second is linear discriminant analysis (LDA) \citep{FISHER1936THE-USE-OF-MULT} which is mostly used for classification tasks. LDA aims to find linear combinations of features that best separate classes of data while maximising the ratio of between-class variance to within-class variance.

Comparisons on previous methods discussed in this paper can be found in \citet{Ghojogh2019Unsupervised-an} and \citet{Chao2019Recent-Advances}. In the former, a detailed theoretical discussion of PCA and SPCA is presented and a comparison between some of the methods presented here can be found. The latter contains a more general review of several supervised dimensionality reduction methods which are not limited to SPCA. The most relevant review for past SPCA methods presented in this paper, including PCR and PLS can be found in \citet{Ritchie2019Supervised-Prin}, where LSPCA is originally presented.

Our proposed method seeks to address two important gaps in the SPCA literature. First, existing approaches struggle to simultaneously maximize (1) relevance of the projected data with respect to the response (covariance with the response variables) and (2) interpretability (variance explained by the projected data). Most methods either prioritize one at the expense of the other. Second, the only method that achieves a trade-off between prediction error and variance explained, LSPCA, relies on a gradient-based manifold optimization scheme, introducing complex mathematical operations and lacking a closed-form solution, which in turn may limit scalability. In this paper, we introduce a novel SPCA framework that addresses both challenges through a mathematically straightforward and elegant, single-optimization formulation. Our method directly balances the trade-off between relevance to the response and explained variance, yielding a closed-form solution via eigenvalue decomposition of a positive semi-definite matrix. This eliminates the need for costly iterative algorithms while ensuring both interpretability and predictive accuracy. To further enhance scalability, we integrate the Nyström approximation, enabling efficient computation even for ultra-high-dimensional data. Our approach provides (1) theoretical rigor via a principled optimization framework with guaranteed convergence, (2) computational efficiency via a simple eigenvalue decomposition and (3) scalability via a low-rank matrix approximation for high dimensional datasets. By unifying relevance and interpretability in a computationally tractable manner, our method advances SPCA beyond heuristic or suboptimal approaches, making it a practical yet theoretically sound choice for modern high-dimensional data analysis.

The rest of the paper is organized as follows. In Section 2, we provide the mathematical background of traditional PCA, existing SPCA methods as well as PLS and LDA which we use as baselines for comparison. In Section 3, we present our proposal for SPCA and its extension using Nyström's method. We examine the performance of our method for prediction and visualization tasks on various simulated and real datasets in Section 4. Finally, Section 5 provides a final discussion and conclusion.

\section{BACKGROUND}\label{back}
In this section, we provide a brief mathematical background for traditional PCA and the methods proposed by \citet{Bairetal}, \citet{Barshan2011Supervised-prin} and \citet{Ritchie2019Supervised-Prin}, along with short descriptions of PLS and LDA for regression and classification, respectively. Let $X\in\mathbb{R}^{n\times p}$ be the data matrix, $Y\in\mathbb{R}^{n\times k}$ the response matrix, $\Sigma=X^\top X$ the covariance matrix and $W\in\mathbb{R}^{p\times q}$ the projection matrix satisfying $W^\top W=I_q$. For binary responses, we will explicitly state so. For the rest of the paper, unless stated otherwise, we assume both $X$ and $Y$ to be centered. Finally let $\|Z\|_F$ denote the Frobenius norm of a matrix $Z$,
\begin{equation*}
\|Z\|_F=\sqrt{\sum_{i}^{m}\sum_{j}^{n}|z_{ij}|^2}=\sqrt{\text{tr}(Z^\top Z)},
\end{equation*}
or equivalently, $\|Z\|_F^2=\text{tr}(Z^\top Z)$.

\subsection{Principal Component Analysis (PCA)}\label{PCA}
PCA seeks a linear projection of the data, $Z=XW$, such that the variance of the projected data is maximized. As an optimization problem, PCA has the two following equivalent formulations,
\begin{enumerate}
    \item[(a)] Minimizing the reconstruction error,
    \begin{equation}\label{PCAsol}
    \min_{W: W^\top W=I_q}\|X-XWW^\top\|_F^2,
    \end{equation}
    \item[(b)] Maximising the variance directly,
    \begin{equation}\label{PCAsol2}
    \max_{W: W^\top W=I_q}\|XW\|_F^2 \: \: := \max_{W:W^\top W=I_q}\text{tr}(W^\top X^\top XW),
    \end{equation}
\end{enumerate}
The projection matrix $W$ consists of the top $q$ eigenvectors of the covariance matrix, $\Sigma$, corresponding to the $q$ largest eigenvalues. These eigenvectors (principal components) can be derived either by performing eigenvalue decomposition on the covariance matrix $\Sigma$, or applying singular value decomposition (SVD) directly on $X$. The process of applying a regression model (linear, logistic etc.) on the projected data, after PCA, is called principal component regression (PCR). 

\subsection{Partial Least Squares (PLS)}
PLS performs a simultaneous decomposition of $X$ and $Y$ such that the covariance of $X$ and $Y$ is maximised. Afterwards, the decomposition of $X$ is used to predict $Y$ in a regression model. The decomposition works as follows,
\begin{equation*}
    X = PU_x^\top  + E_x 
\end{equation*}
\begin{equation*}
    Y = KU_y^\top  + E_y,
\end{equation*}
where $P,K\in\mathbb{R}^{n\times q}$, for the $q$ extracted components known as latent vectors, $U_x\in\mathbb{R}^{p\times q}$ and $U_y\in\mathbb{R}^{k\times q}$ are the loading matrices and $E_x$ and $E_y$ are the residual matrices of dimensions $n\times p$ and $n\times k$ respectively. As an optimization problem, PLS can be defined as,
\begin{equation}\label{PLS}
    \max_{r,s:|r|=|s|=1}\text{cov}(Xr,Ys),
\end{equation}
where $r$ and $s$ are weight vectors. In order to solve (\ref{PLS}), one usually needs to perform an iterative algorithm which we do not present here, but can be found in \citet{Rosipal2006Overview-and-Re}.
PLS differs from all other methods we present here in that it also decomposes the response matrix as well as the data matrix. 

\subsection{Linear Discriminant Analysis (LDA)}
LDA seeks to find an optimal projection that maximizes class separability by minimizing within-class variance while maximizing between-class variance \citep{Zhao2024Linear-discrimi}. Unlike PCA, which seeks to find the direction where the variance is maximized, LDA explicitly finds the direction for classification performance. Consider a binary response $Y\in\{0,1\}^n$ defining two classes, $C_0$ and $C_1$, with $n_c$ samples in each class ($c=0,1$). Let $M_c=n_c^{-1} \sum_{j=1}^{n_c}X_j$ be the mean vector for class $c$ and $M=n^{-1}\sum_{i=1}^{n}X_i$ the overall mean. 
The between-class scatter matrix measures class separation,
\[S_b=\sum_{c=0}^{1}n_c(M_c-M)(M_c-M)^\top,\]
while the within-class scatter matrix measures intra-class variability,
\[S_w=\sum_{c=0}^{1}\sum_{j\in C_c}(X_j-M_c)(X_j-M_c)^\top.\]
LDA finds a projection matrix $W$ that maximizes Fisher's criterion,
\[\arg\max_{W}\frac{W^\top S_bW}{W^\top S_wW}.\]
This is equivalent to solving the generalized eigenvalue problem 
\[S_w^{-1}S_bW=\Lambda W,\]
where $\Lambda$ is a diagonal matrix of eigenvalues. The optimal projection consists of the top $q$ eigenvectors of $S_w^{-1}S_b$ corresponding to the $q$ laregst eigenvalues.

\subsection{Bair's Method}\label{Bair}
SPCA was first introduced in \citet{Bair2004Semi-Supervised} and \citet{Bairetal}, referred to here as Bair's method. Unlike recent SPCA methods that directly integrate the response into dimensionality reduction, Bair's method operates in two stages, relying solely on the feature data for projection,
\begin{enumerate}[I.]
    \item Fit univariate regression models for each feature $X_j$ (e.g. linear, logistic etc.) and compute the standard regression coefficients,
    \begin{equation*}
        \beta_j = \frac{X_{j.}Y^\top}{\sqrt{X_{j.}X_{j.}}},
    \end{equation*}
    where $X_{j.}$ denotes the $j$-th row of $X$. Form the matrix $X_{\theta}$ consisting of features whose coefficients exceed the threshold value $\theta$ in absolute value, i.e. $|\beta_j|>\theta$.
    \item Apply PCA to $X_{\theta}$ to obtain projections $Z_{\theta}=X_{\theta}W_{\theta}$, where $W_{\theta}$ is the orthogonal loading matrix and fit a regression model on $Z_{\theta}$ for prediction.
\end{enumerate}

The value of $\theta$ in Step I is calculated using cross-validation. The primary limitations of this method are (1) feature selection bias, since it can discard features that contribute jointly but not individually, (2) it is restricted for a single response variable and (3) it remains unclear how the selection of the threshold value $\theta$ affects the trade-off between prediction and variance explained.

\subsection{SPCA using HSIC}\label{HSIC}
\citet{Barshan2011Supervised-prin} formulate SPCA through the Hilbert-Schmidt independence criterion (HSIC), referred to here as SPCA using HSIC. This states that in a universal reproducing kernel Hilbert space (RKHS), two variables are independent if and only if their HSIC is equal to $0$. Using an empirical measure of HSIC the optimization problem can be formualted as,
\begin{equation}\label{Barshan}
    \max_{W:W^\top W=I_q}\text{tr}(W^\top X K X^\top W),
\end{equation}
where $W^\top XX^\top W$ is a kernel of $XW$ and $K$ is a kernel of $Y$. For regression tasks, it can be defined as the  radial basis function (RBF) kernel,  
$$K=K(y,y')=\exp{\frac{-\|y-y'\|^2}{2\sigma^2}},$$
while for classification tasks, the delta kernel is the most common choice, \[\delta(y,y')=\begin{cases} 
1 & \text{if } y = y', \\
0 & \text{if } y \neq y'.
\end{cases}\]

The main benefit of SPCA using HSIC is that it offers a closed-form solution through eigenvalue decomposition of $Q=XKX^\top$, by taking the top $q$ eigenvectors corresponding to the top $q$ largest eigenvalues of $Q$. However, the objective function does not include a variance term resulting in the projected data not fully capturing the intrinsic variability of the original data as PCA would do. Additionally, the performance depends on appropriate choice of the kernel hyperparameters (e.g. $\sigma$ in the RBF kernel).

Recently, \citet{Pascual2022Least-squares-r} proposed least squares regression principal component analysis, that shares the exact same objective function as SPCA using HSIC but impose a different constraint on the optimization problem. Specifically, instead of the usual orthogonality condition on the projection matrix, that is $W^\top W=I_q$, they impose the orthogonality constraint on $XW$, i.e. $W^\top X^\top XW=I_q$. Due to the identical objective function of the two methods, we focus on SPCA using HSIC due to its wider recognition among SPCA methods.

\subsection{Least Squares PCA}\label{LSPCA}
Least squares PCA (LSPCA) \citep{Ritchie2019Supervised-Prin,Ritchie2020Supervised-PCA:} reformulates SPCA as a manifold optimization problem, balancing dimensionality reduction and prediction via a tunable trade-off parameter $\lambda>0$. The objective combines 
\begin{itemize}
    \item An arbitrary supervised loss $L(Y,XW,\beta)$ (e.g. MSE for regression, logistic loss for classification),
    \item An unsupervised PCA term $\|X-XWW^\top\|_F^2$ (reconstruction error).
\end{itemize}
The optimization problem is constrained on the Stiefel manifold, i.e. the set of all matrices with orthonormal columns ($W^\top W=I_q$),
\begin{equation}\label{LSPCAobj}
    \min_{\beta,W}L(Y,XW,\beta)+\lambda\|X-XWW^\top\|_F^2,
\end{equation}
The main advantage of this method is that it simultaneously learns a projection $W$ and the regression coefficients $\beta$, leading to direct predictions via $\hat{Y}=X_{\text{new}}W^\top\hat{\beta}$. However, a primary drawback of this method is that it does not offer a closed-form solution. Instead, it requires gradient descent on the Grassmannian manifold, complicating implementation. Additionally, we find that the algorithm is sensitive to the choice of the hyperparameter $\lambda$, as small changes in $\lambda$ significantly affect performance, making tuning more complicated, especially in very high dimensional settings.

\section{Covariance Supervised PCA (CSPCA)}
In this section, we propose a novel SPCA method that seeks to project the data onto a lower dimensional space such that the covariance between the projected data and the response variables is maximised while also preserving as much of the variance explained by the original data similar to PCA. In order to balance these two competing objectives, we introduce a regularisation parameter to control the two terms depending the task at hand. We call this method covariance supervised principal component analysis (CSPCA). We also provide a computationally efficient extension of CSPCA to account for very large values of $p$. This section is organized as follows. First, we provide the motivation and the limitations in existing SPCA literature that CSPCA aims to address. Second, we provide the mathematical framework of CSPCA that derives the projection matrix from a simple eigenvalue decomposition. Next, we provide an extension of CSPCA using Nyström’s approximation for scalable application in high dimensional settings, e.g. microarray data. Finally, we provide a discussion on the computational complexity of CSPCA and its extension using Nyström’s method.

\subsection{Motivation}
The method we describe aims to address current limitations in the SPCA literature, while also providing an interpretable, robust and scalable framework. One of the primary limitations in SPCA methods, is that they do not address the simultaneous objective of maximising information between the response and the projected data while also preserving the intrinsic variability of the original data. As a result, in practice, SPCA methods tend to favor performance in one aspect. For example, SPCA using HSIC \citep{Barshan2011Supervised-prin} seeks to maximize the dependence between the response and the projected data, potentially improving predictive performance, but does not address the variance explained by the projected data, leading to less interpretable projections. The only method balancing the two objectives, LSPCA \citep{Ritchie2019Supervised-Prin}, relies on computationally intensive manifold gradient optimization, which is also sensitive to hyperparameter tuning and lacks a closed-form solution. We propose a novel method, CSPCA, which seeks to overcome the two gaps of the SPCA literature. Specifically, CSPCA defines an optimization framework that incorporates an unsupervised term to address the variance explained by the projected data as in PCA and a supervised term to improve prediction performance and preserve the overall information between the response and the projected data. Since the unsupervised and supervised objectives are competing each other, we introduce a hyperparameter, $\kappa$, to balance between the competing objectives. In contrast to LSPCA, CSPCA is not sensitive to the choice of the balancing hyperparameter. This stems mostly from the fact that CSPCA derives the projection through a fixed closed-form solution, while LSPCA relies on an optimization algorithm which can be affected by the choice of the parameter at every run. In the following subsection, we provide the mathematical formulation of our proposed method.

\subsection{Theoretical Foundations of CSPCA}
The framework of CSPCA, as the name suggests, is based on covariance, hence continuous responses and is later adjusted accordingly for binary responses and classification tasks. Assuming both the data features, $X$, and the response variables, $Y\in\mathbb{R}^{n\times k}$, have already been centered, we can define the cross-covariance between the projected data $Z=XW$ and the responses $Y$ as,
\begin{equation*}
    \text{cov}(Z,Y)=\text{cov}(XW,Y)=\frac{1}{n}(XW)^\top Y=\frac{1}{n}W^\top X^\top Y.
\end{equation*}
The overall covariance between the data and the response variables is defined as the sum of all the  elements of the cross-covariance matrix,
\begin{equation}\label{Cov}
    f_\text{\rm Cov} = \sum_{i=1}^{q}\sum_{j=1}^{k}\text{cov}(Z_i,Y_j).
\end{equation}
Instead of maximising this quantity directly, we take the squared Frobenius norm of the cross-covariance matrix,
\begin{equation}\label{CovNew}
    \|\text{cov}(Z,Y)\|_F^2=\|W^\top X^\top Y\|_F^2=\text{tr}(Y^\top XWW^\top X^\top Y),
\end{equation}
where we have dropped the scalar term $n^{-2}$. The Frobenius norm provides a measure of the overall strength of the relationship between the projected data and the response variables. The objective function in (\ref{CovNew}) defines the supervised part of our overall objective, $\|W^\top X^\top Y\|_F^2$.

The unsupervised objective follows the idea of PCA, which seeks to maximize the variance of the projected data $Z=XW$. The covariance matrix of the projected data is defined as
\begin{equation*}
    \text{cov}(Z)=\frac{1}{n}Z^\top Z=\frac{1}{n}W^\top X^\top XW.
\end{equation*}
The overall variance of the projected data is given by the sum of the diagonal elements of the covariance matrix of the projected data which is equal to the square of the Frobenius norm of the projected data matrix, 
\begin{equation}\label{Var}
f_\text{\rm Var}=\|XW\|_F^2=\text{tr}(W^\top X^\top XW).
\end{equation}
In contrast to the majority of PCA applications that use the reconstruction error formulation (\ref{PCAsol}), we herein adopt directly the idea of maximising the variance of the projected data (\ref{Var}). This introduces the unsupervised objective of our overall objective function, $\|XW\|_F^2$.

%As we have already mentioned, 
The supervised and unsuperivsed objectives are competing. For example, when maximising the covariance between the response and the projected data, this does not necessarily lead to a projection that captures the intrinsic variability of the data $X$, which can also be of interest after projecting the data into lower dimensions. Similarly, when maximising the variance of the projected data, the resulting projection is not guaranteed to be informative of the response variables.

To overcome this, we introduce a regularisation parameter, $\kappa>0$, to balance the two competing objectives. The parameter is imposed on the unsupervised term and the proposed optimization problem is defined as 
\begin{equation}{\label{Eq1}}
    \max_{W: W^\top W=I_q}\{\|W^\top X^\top Y\|^2_F+\kappa\|XW\|^2_F\},
\end{equation}
or, equivalently,
\begin{equation}{\label{Eq2}}
    \max_{W: W^\top W=I_q}\{\text{tr}(Y^\top XWW^\top X^\top Y)+\kappa \text{tr}(W^\top X^\top XW)\}.
\end{equation}

The choice of the hyperparameter $\kappa$ is task-dependent and should be determined according to the specific objectives of the study. If the primary objective is prediction performance, we propose a cross validation (CV) scheme to tune $\kappa$ in Algorithm \ref{alg:K-Fold}. In general, larger values of $\kappa$ put more weight on the unsupervised objective and lead to better performance in terms of variance explained by the projected data. Similarly, smaller values of $\kappa$ lead the supervised objective to dominate (\ref{Eq1}), resulting in more informative projections with respect to the response variables, while still preserving a substantial amount of the variance of the original data. However, it is important to note that a small value of $\kappa$ does not necessarily translate to better prediction performance and that is why we suggest a CV scheme for tuning $\kappa$ when the goal is for better prediction performance. The detailed steps to perform $K$-fold CV to tune $\kappa$ are presented in Algorithm \ref{alg:K-Fold}.

The primary strength of CSPCA is that it offers a closed-form solution via eigenvalue decomposition. This can be achieved by rewriting the objective function using properties of matrix traces. From (\ref{Eq2}), the unsupervised objective is $W^\top X^\top XW\in\mathbb{R}^{q\times q}$ and the supervised objective is $Y^\top XWW^\top X^\top Y\in\mathbb{R}^{k\times k}$. 
Using properties of the trace, we have
\begin{eqnarray}
&&\text{tr}(Y^\top XWW^\top X^\top Y)+\kappa\text{tr}(W^\top X^\top XW)\nonumber\\
&=&\text{tr}((Y^\top XW)(W^\top X^\top Y))+\text{tr}(\kappa W^\top X\top XW). \label{A}
\end{eqnarray}
Since $Y^\top XW\in\mathbb{R}^{k\times q}$ and $W^\top X^\top Y\in\mathbb{R}^{q\times k}$, we can write 
\begin{equation*}
    \text{tr}((Y^\top XW)(W^\top X^\top Y))=\text{tr}(W^\top X^\top YY^\top XW)
\end{equation*}
and using the linear combination property of the trace in (\ref{A}), we can simplify the objective function as
\begin{eqnarray*}
&&\text{tr}(W^\top X^\top YY^\top XW) + \text{tr}(\kappa W^\top X^\top XW)\nonumber\\
%&=&\text{tr}(W^\top X^\top YY^\top XW+\kappa W^\top X^\top XW)\nonumber\\
&=&\text{tr}(W^\top(X^\top YY^\top X+\kappa X^\top X)W)=\text{tr}(W^\top CW),
\end{eqnarray*}
where $C=X^\top YY^\top X+\kappa X^\top X\in\mathbb{R}^{p\times p}$. The matrix $C$ is symmetric and positive semi-definite as the sum of symmetric and positive semi-definite matrices. Thus, the proposed optimization problem in (\ref{Eq2}) can be reduced to \begin{equation}\label{Eq3}
     \max_{W: W^\top W=I_q}\{\text{tr}(W^\top CW)\},
\end{equation}
where $W^\top CW\in\mathbb{R}^{q\times q}$.

The solution to (\ref{Eq3}) can easily be derived, similarly to PCA, by taking the eigenvalue decomposition of $C$, 
\begin{equation}\label{C1}
    C=U\Lambda U^{-1},
\end{equation}
where $\Lambda=\text{diag}\{\lambda_1,\ldots,\lambda_p\}$ is the diagonal matrix containing the eigenvalues of $C$ and $U$ is the $p\times p$ matrix of eigenvectors of $C$. The projection matrix $W$ is defined as the matrix consisting of the top $q$ eigenvectors of $C$ that correspond to the $q$ largest eigenvalues, i.e. $W=[u_1,\ldots,u_q]$. After computing $W$, we can project the data into the lower dimensional space as $Z=XW$, where $Z\in\mathbb{R}^{n\times q}$. The projected data can then be used for visualization or applying predictive models. 

The proposed framework assumes continuous response variables and models their relationship with the features through covariance. However, covariance is inherently designed for continuous variables and fails to capture the discrete nature of binary outcomes, making the original formulation of CSPCA suboptimal for classification tasks. Instead, we use a delta kernel to model the similarities between the binary labels,
\begin{equation}
    \Delta = \delta(y,y')=\begin{cases}
        1, \: \: \text{if} \: \: y=y' \\
        0, \: \: \text{if} \: \: y\neq y'
    \end{cases}
\end{equation}
where $\Delta$ is used to highlight that the kernel refers specifically to the delta kernel.

We integrate the delta kernel in our objective function by swapping $YY^\top$ in the supervised part of $C$, $X^\top YY^\top X$, with $\Delta$ to derive $X^\top \Delta X$. There are several benefits for using the delta kernel in this situation. First of all, the delta kernel encodes whether two samples belong to the same class, improving the interpretability of the objective. Additionally, it preserves the positive semi-definiteness of $C$. Finally, it does not assume a linear relationship between the response variables and the data, rather it focuses on grouping similar responses together, extending CSPCA's use for non-linear settings. In its complete form, $C$ can now be written as
\begin{equation}
    C= X^\top \Delta X + \kappa X^\top X.
\end{equation}
The optimization problem in (\ref{Eq3}) remains unchanged, as well as its solution via eigenvalue decomposition of $C$. 

The mathematical framework we have presented here is simple yet robust and provides a straightforward approach to integrate a supervised and an unsupervised term in a single optimization framework while still offering a closed-form solution. The introduction of the balancing hyperparameter provides a wide spectrum of solutions, providing simultaneously interpretable and relevant projections, not achievable by single objective methods like PLS or SPCA using HSIC. In the next section, we introduce a scalable extension of CSPCA using low-rank matrix approximation for cases where the number of features is substantially large, e.g. thousands to tens of thousands, in order to reduce computational burden.

\begin{algorithm}[h]
\caption{$K$-Fold Cross-Validation for Tuning \(\kappa\) in CSPCA}
\label{alg:K-Fold}
\begin{algorithmic}[1]
\State \textbf{Input:} 
\State \quad \(X\in\mathbb{R}^{n\times p}\), \(Y\in\mathbb{R}^{n\times k}\) 
\State \quad \(\mathcal{K}_{\text{values}}\): list of candidate \(\kappa\) values
\State \quad \(q\): number of components
\State \quad \(K\): number of folds for cross-validation
\State \textbf{Output:} 
\State \quad \(\hat{\kappa}\): optimal \(\kappa\) value
\State \quad \(\text{best\_score}\): best average validation MSE
\State Initialize \(\text{best\_score} \gets \infty\), \(\hat{\kappa} \gets \text{None}\)
\State Split data \((X, Y)\) into \(K\) folds \(\{(X_k,Y_k)\}_{k=1}^{K}\)
\For{each \(\kappa \in \mathcal{K}_{\text{values}}\)}
    \State Initialize \(\text{fold\_scores} \gets []\)
    \For{each fold \(k \in \{1, \dots, K\}\)}
        \State Define validation set:  \((X_{\text{val}}, Y_{\text{val}}) \gets (X_k,Y_k)\)
        \State Define training set: \((X_{\text{train}}, Y_{\text{train}})\gets \cup_{j\neq k}(X_j,Y_j)\)
        \State Compute \(C \gets X_{\text{train}}^\top Y_{\text{train}} Y_{\text{train}}^\top X_{\text{train}} + \kappa X_{\text{train}}^\top X_{\text{train}}\)
        \State Compute \(W\) as the top \(q\) eigenvectors of \(C\) 
        \State Project training data: \(Z_{\text{train}} \gets X_{\text{train}} W\)
        \State Project validation data: \(Z_{\text{val}} \gets X_{\text{val}} W\)
        \State Fit a regression model $\mathcal{M}$ on \((Z_{\text{train}}, Y_{\text{train}})\)
        \State Predict \(Y_{\text{pred}}\) \(\gets\mathcal{M}(Z_{\text{val}})\)
        \State Compute \(\text{MSE}_k\gets\frac{1}{n_{\text{val}}}\|Y_{\text{val}}-Y_{\text{pred}}\|^2_F\)
        \State Append \(\text{MSE}_k\) score to \(\text{fold\_scores}\)
    \EndFor
    \State Compute \(\text{avg\_mse} \gets \text{mean}(\text{fold\_scores})\)
    \If{\(\text{avg\_mse} < \text{best\_score}\)}
        \State \(\text{best\_score} \gets \text{avg\_mse}\)
        \State \(\hat{\kappa} \gets \kappa\)
    \EndIf
\EndFor
\State \Return \(\hat{\kappa}\), \(\text{best\_score}\)
\end{algorithmic}
\end{algorithm}

\subsection{Computationally Scalable CSPCA}
CSPCA effectively projects high-dimensional data into a lower-dimensional subspace by simultaneously maximizing relevance to the response variables and preserving interpretability through explained variance. It is particularly well-suited for high-dimensional settings where the number of features is much greater than the number of samples, i.e. $p\gg n$. However, when $p$ becomes extremely large, as is common in microarray experiments where $p$ ranges from $10^3$ to $10^4$, the computational cost of performing eigendecomposition on the $p\times p$ objective matrix becomes prohibitive. This limitation motivates the development of a scalable approximation to CSPCA that retains its statistical properties while significantly reducing computational burden. 

To address this challenge, we employ Nyström's method \citep{Nystrom1930Uber-Die-Prakti}, a well-established technique for low-rank approximation of large symmetric positive semi-definite matrices. The Nyström approximation has been widely adopted in machine learning for scaling kernel-based methods such as PCA \citep{Arcolano2011Estimating-prin}, kernel PCA \citep{Sterge2022Statistical-opt} and kernel Fisher discriminant analysis \citep{Wang2013A-novel-multipl}. By constructing a low-rank surrogate of the original matrix, Nyström’s method reduces the computational complexity of eigendecomposition from $\mathcal{O}(p^3)$ to $\mathcal{O}(m^3)$, where $m\ll p$ is the number of sampled landmark points.

Nyström's method works by sampling a subset of columns from a symmetric positive semi-definite matrix and then uses the correlations between the sampled columns and the remaining ones to form a low-rank approximation of the original full matrix. \citet{Sun2015A-review-of-Nys} described how Nyström's method can be divided into three steps. First, a sampling step takes place, where a subset of $m$ columns is uniformly sampled from the original $p\times p$ matrix, forming a submatrix. A singular value decomposition step is then performed on the intersection of the sampled columns and their corresponding rows, yielding singular values and vectors. For the final extension step, the partial eigendecomposition is extrapolated to approximate the full set of eigenvectors and eigenvalues of the original matrix.

This approach significantly reduces the computational complexity of the original eigenvalue decomposition problem, making it an attractive alternative for problems with large matrices. Although advanced sampling strategies \citep{Drineas2005On-the-Nystrom-,Kumar2009Ensemble-Nystro} can further optimize the approximation quality, we adopt uniform sampling for its simplicity and empirical effectiveness as it is shown to perform well in practice.

We follow the notation introduced in the previous section and directly describe how Nyström's method can be tailored to the CSPCA optimization problem. Assume that $C$ is a symmetric positive semi-definite matrix such that $C\in\mathbb{R}^{p\times p}$. We randomly sample $m$ columns from $C$ and let $S$ denote the matrix containing the sampled columns, i.e. $S\in\mathbb{R}^{p\times m}$. The columns of $S$ and $C$ can be rearranged so that
\[
S = \begin{bmatrix} 
C_m \\ D 
\end{bmatrix} 
\quad \text{and} \quad
C = \begin{bmatrix} 
C_m \ \ D^\top\\
D \ \ B 
\end{bmatrix},
\]
where $C_m\in\mathbb{R}^{m\times m}$ is the submatrix containing the intersection of the sampled columns and rows, $D\in\mathbb{R}^{(p-m)\times m}$ and $B\in\mathbb{R}^{(p-m)\times(p-m)}$. Assume that the SVD of $C_m$ is
\begin{equation*}
  C_m=U_m\Lambda_mU_m^\top,  
\end{equation*}
where $U_m$ is an orthonormal matrix, $\Lambda_m=\text{diag}(\sigma_1,\dots,\sigma_m)$, and $\sigma_1,\dots,\sigma_m$ are the singular values of $C_m$. The Nyström approximation of $C$ can then be calculated as
\begin{equation}\label{C}
    C\approx\tilde{C}=SC_m^+S^\top, 
\end{equation}
where $C_m^+$ is the Moore-Penrose pseudoinverse matrix of $C_m$ \citep{Li2010Making-large-sc}. We can thus use (\ref{C}) to approximate the eigenvectors of $C$ as
\begin{equation}\label{ext}
    U\approx SU_m\Lambda_m^{-1/2}.
\end{equation}
The approximation in (\ref{ext}) is usually referred to as Nyström's extension \citep{Sun2015A-review-of-Nys}. 

Connecting back to CSPCA, we can derive the projection matrix $W$ from (\ref{Eq3}) by taking the top $q$ eigenvectors of the approximated matrix of eigenvectors $U$.

The rank of Nyström's approximation, $\tilde{C}$, is at most $m$. This implies that the quality of the approximation is heavily influenced by the choice of $m$. If $m$ is too small, the approximation might be poor, while if $m$ is large, the computational benefits of the method diminish, due to the cubic dependence on $m$. In practice, $m$ is chosen as a trade-off between accuracy and computational cost, with popular choices being $m=\sqrt{p}$ or $p^{1/3}$.

\begin{table*}[t]
\centering
\caption{Computational Complexities of CSPCA and Extended CSPCA}
\label{tab:complexities}
\resizebox{\textwidth}{!}{%
\begin{tabular}{lll}
\toprule
\textbf{Method} & \textbf{Total Complexity} & \textbf{\hspace{.5in}Individual Complexity} \\
\midrule
\textbf{CSPCA} &
\begin{tabular}{@{}l@{}}
\( \mathcal{O}(np^2 + npk + p^3) \) \\
 (\( \approx \mathcal{O}(p^3) \) for large $p$)
\end{tabular} & 
\begin{minipage}{0.5\textwidth}
\begin{itemize}
    \item[] \( \mathcal{O}(np^2) \): Matrix multiplication of \( X^\top X \) 
    \item[] \( \mathcal{O}(npk) \): Matrix multiplication of \( X^\top Y \) and \( Y^\top X \)
    \item[] \( \mathcal{O}(p^3) \): Eigenvalue decomposition of \( C \)
\end{itemize}
\end{minipage} \\
\midrule
\textbf{Extended CSPCA} & 
\begin{tabular}{@{}l@{}}
\( \mathcal{O}(pm^2 + p^2 m + m^3) \) \\
(\( \approx \mathcal{O}(p^2 m) \) when $m\ll p$)
\end{tabular} & 
\begin{minipage}{0.6\textwidth}
\begin{itemize}
    \item[] \( \mathcal{O}(pm^2) \): Nyström extension
    \item[] \(\mathcal{O}(p^2 m)\):  Nyström approximation \( S C_m^+S^\top\)
    \item[] \( \mathcal{O}(m^3) \): SVD of \( C_m \) and $C_m^+$ computation
\end{itemize}
\end{minipage} \\
\bottomrule
\end{tabular}
}
\end{table*}

\subsection{Computational Complexity of CSPCA}\label{CompComp}
The computational cost of CSPCA is dominated by its eigenvalue decomposition step, which scales cubically with the number of features, $\mathcal{O}(p^3)$. As detailed in Table \ref{tab:complexities}, the total computational complexity of CSPCA is $\mathcal{O}(pnk +np^2+p^3),$
where $p$ is the number of features, $n$ is the number of samples and $k$ is the number of response variables. In most cases $k \ll p$ and the dominant term is $\mathcal{O}(np^2+p^3)$. In high dimensional settings, such as microarray experiments, where the number of features ranges from thousands to tens of thousands, the cubic term $\mathcal{O}(p^3)$ becomes prohibitively expensive, making CSPCA computationally intractable for large-scale problems.

To mitigate this limitation, we have employed Nyström's approximation which significantly reduces the cost of the eigenvalue decomposition from $\mathcal{O}(p^3)$ to $\mathcal{O}(m^3)$, where $m\ll p$ is the number of sampled columns. The total computational complexity of the Nyström-extended CSPCA is then $\mathcal{O}(m^3+pm^2+p^2m).$ A common heuristic choice is to select $m$ as $\sqrt{p}$ or $p^{1/3}$, substantially reducing the dominant term to $\mathcal{O}(p^2m)$. Since $p \gg m$, this results in an effective complexity of $\mathcal{O}(p^2)$, a significant improvement over the original $\mathcal{O}(p^3)$. A detailed breakdown of these computational costs is provided in Table \ref{tab:complexities}. In practice, we find that for a dataset with more than $p=7000$ features, Nyström-extended CSPCA requires only a few seconds to run. In summary, the Nyström extension transforms the computational bottleneck of CSPCA from a cubic dependence on $p$ to a quadratic one, enabling scalable application to high-dimensional datasets without sacrificing numerical stability or statistical performance.

\section{EXPERIMENTAL RESULTS}
We conduct a series of simulation analyses along with applications on four real datasets in order to examine CSPCA's performance in practice compared to the state-of-the-art SPCA methods and baselines. All datasets used are publicly available and we also provide software implementation for CSPCA and our analyses.

\subsection{Simulations}
We conduct a comprehensive simulation study to evaluate the performance of the SPCA methods in high-dimensional regression settings, where the number of predictors $p=600$ exceeds the sample size $n=100$. We consider a linear and two non-linear data-generating processes (DGP) for the response variable $Y$ as follows:
\begin{enumerate}
    \item $Y=3X_1 - 5X_2+4X_3+\epsilon$, with $\epsilon\sim N(0,0.1^2)$, \newline
    \item $Y=e^{X_1}+3\sin X_2-4X_3+\epsilon$, with $\epsilon\sim N(0,0.1^2)$, \newline
    \item $Y=2X_1-\frac{1}{2}X_1^2+3X_2+4e^{(X_3+1)}+\epsilon$, with $\epsilon\sim N(0,0.1^2)$.
\end{enumerate}

For the design matrix $X$, we consider two distinct covariance structures. First, we assume $X\sim N(0,I_p)$, where features are independent and identically distributed (i.i.d.) standard normal variables. Next, we introduce correlated features, $X\sim N(0,\Sigma)$, with a Toeplitz covariance matrix $\Sigma$ defined by $\Sigma_{ij}=\rho^{|i-j|}$. We set $\rho=0.4$, inducing moderate correlation between features, such that the correlation between $X_i$ and $X_j$ decays exponentially with $|i-j|$. 

For each scenario, we generate 50 independent datasets, each split into $80\%$ training and $20\%$ test sets. Performance is assessed using three key metrics: variance explained, mean squared error (MSE) and covariance explained. Using the notation introduced in Section \ref{back}, we define variance explained by the projected data as the ratio of the variance of the projected data over the variance of the original data,  
${\|XW\|^2_F}/{\|X\|^2_F}.
$
The MSE is defined as
${n}^{-1}\|\hat{Y}-Y\|^2_F,
$
where $\hat{Y}$ corresponds to the predicted values of the response variable and the covariance explained as the ratio of the Frobenius norm of the cross-covariance matrix between the projected data and the response variables over the Frobenius norm of the cross-covariance matrix between the original data and the response variables,
$
    {\|W^\top X^\top Y\|^2_F}/{\|X^\top Y\|^2_F}.
$
The latter serves as an indicator of how informative the projected data are to the response variables.

\subsubsection{Simulation 1}
We begin by analyzing the scenario where the response variable $Y$ is linearly associated with the predictors. Figure \ref{fig:iid} presents the results for the i.i.d. setting (top three plots), while Figure \ref{fig:cor} displays the corresponding results for the correlated scenario (top three plots). 

In the i.i.d. case, CSPCA and LSPCA exhibit behaviors similar to PCA in terms of variance explained, particularly as the number of components $q$ increases, where their performance becomes nearly indistinguishable for larger $q$, with CSPCA even surpassing PCA for $q=8$ and $10$ components. In terms of MSE, CSPCA along with PLS and LSPCA emerge as the top-performing methods, clearly outperforming other methods across all numbers of components ($q$). 

CSPCA demonstrates a clear advantage in recovering covariance structures relevant to the response, with LSPCA providing competitive but slightly inferior performance. In contrast, PCA and SPCA using HSIC perform poorly in terms of both prediction error and covariance explained, while Bair’s method performs better, but as expected, yields the lowest variance explained due to its restrictive assumptions. Overall, CSPCA strikes an effective balance across all performance metrics, achieving the best performance through a computationally efficient closed-form solution, unlike LSPCA, which relies on a more complex gradient-based optimization framework.

Similar conclusions can be made for the correlated scenario regarding variance and covariance explained. Both CSPCA and LSPCA perform close to PCA in variance explained, approaching PCA's performance as the number of components increases. However, CSPCA maintains an advantage over LSPCA in covariance explained, while the other methods underperform substantially in both metrics. The most notable difference, however, is observed in terms of prediction error where CSPCA remains consistent with its i.i.d. scenario performance, while PLS and LSPCA get worse as the number of components increases. Consequently, CSPCA emerges as the best-performing method in terms of MSE in this case, establishing its strong performance across all performance metrics.

\subsubsection{Simulation 2}
We next examine the scenario where the response variable $Y$ is nonlinearly associated with the predictors, incorporating exponential and trigonometric terms. Figure \ref{fig:iid} presents results for the i.i.d. setting (middle three plots), while the correlated feature scenario is shown in Figure \ref{fig:cor} (middle three plots).

For the i.i.d. case, results do not differ significantly from the linear simulation. Once more, LSPCA and CSPCA challenge PCA's performance in terms of variance explained, with CSPCA even surpassing it for $q=8$ and $10$ components. Regarding prediction error, PLS, LSPCA and CSPCA significantly outperform remaining methods, however, CSPCA provides the lowest MSE across all components. Finally, CSPCA and LSPCA are the top-performing methods in terms of preserving relevant structures with the response, clearly outperforming remaining methods. Bair's method also illustrates strong performance in terms of covariance explained, however it is diminished by its worse performance in terms of variance explained. It is also worth noting that PCR which is the only method that does not incorporate the response variable in its framework, is the worst performing method in terms of prediction error and covariance explained.

Identical conclusions can be made for the correlated scenario, especially in terms of variance and covariance explained, where LSPCA and CSPCA are the top-performing methods, approaching PCA's level of variance explained. The most notable difference is again observed for prediction error. Specifically, in this case, CSPCA's superiority over PLS and LSPCA in terms of MSE is more evident compared to the i.i.d. case as can be seen by \ref{Fig:2e}. In conclusion, we observe that the addition of non-linear terms, such as exponential or trigonometric do not affect the performance of CSPCA.

\subsubsection{Simulation 3}
\begin{figure}[h]
    \centering
    % Row 1
    \begin{subfigure}{0.32\textwidth}
        \centering
        \includegraphics[width=\linewidth]{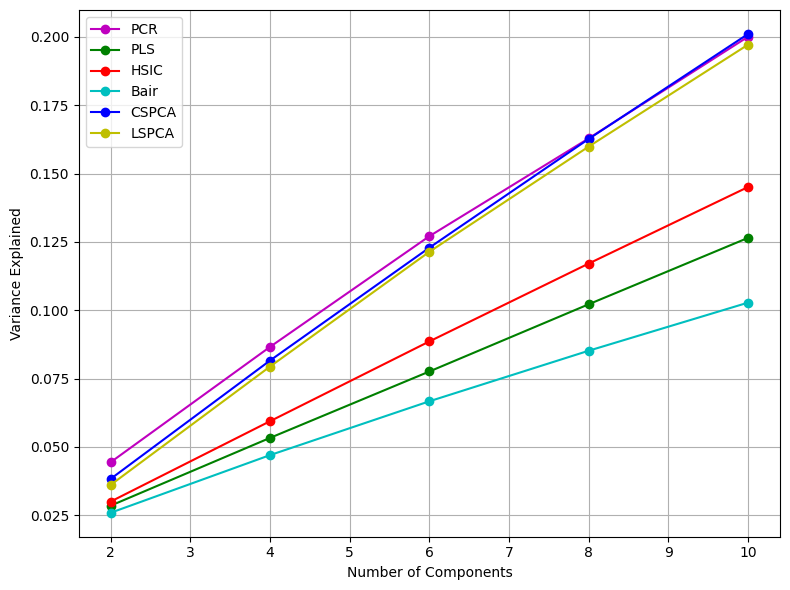}
        \caption{Variance Explained (Sim 1)}
        \label{Fig:1a}
    \end{subfigure}
    \hfill
    \begin{subfigure}{0.32\textwidth}
        \centering
        \includegraphics[width=\linewidth]{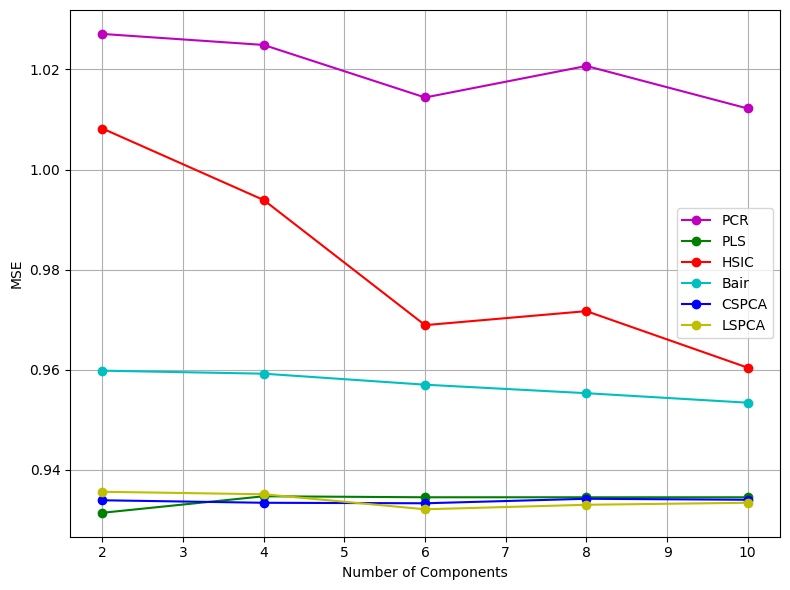}
        \caption{Mean Squared Error (Sim 1)}
        \label{Fig:1b}
    \end{subfigure}
    \hfill
    \begin{subfigure}{0.32\textwidth}
        \centering
        \includegraphics[width=\linewidth]{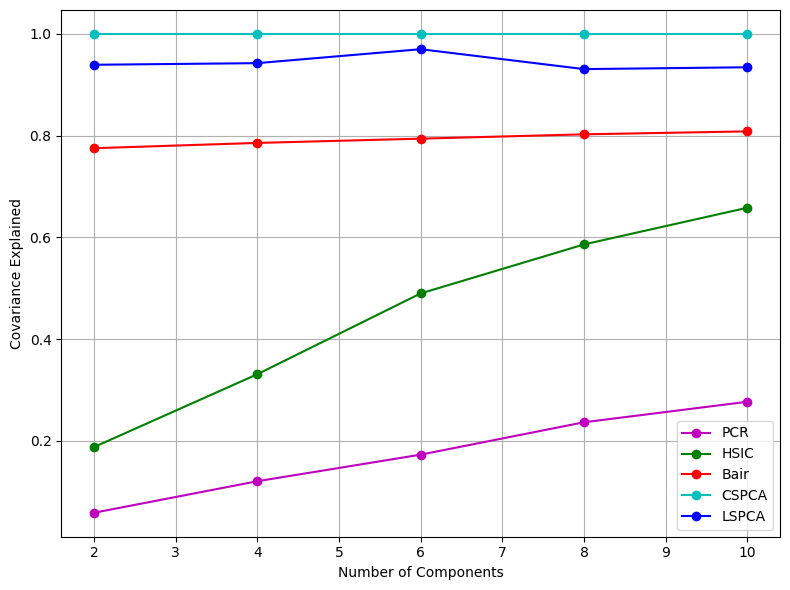}
        \caption{Covariance Explained (Sim 1)}
        \label{Fig:1c}
    \end{subfigure}

    \vspace{1em} % Add vertical spacing between rows

    % Row 2
    \begin{subfigure}{0.32\textwidth}
        \centering
        \includegraphics[width=\linewidth]{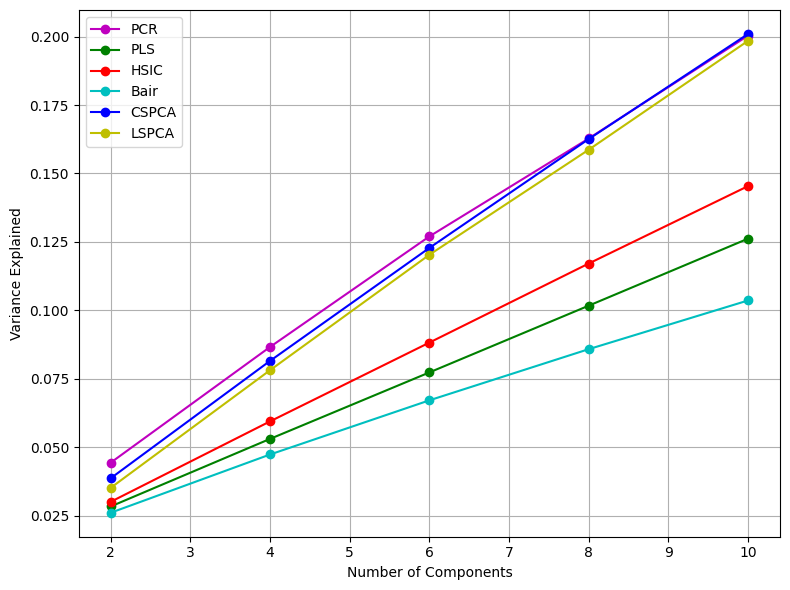}
        \caption{Variance Explained (Sim 2)}
        \label{Fig:2a}
    \end{subfigure}
    \hfill
    \begin{subfigure}{0.32\textwidth}
        \centering
        \includegraphics[width=\linewidth]{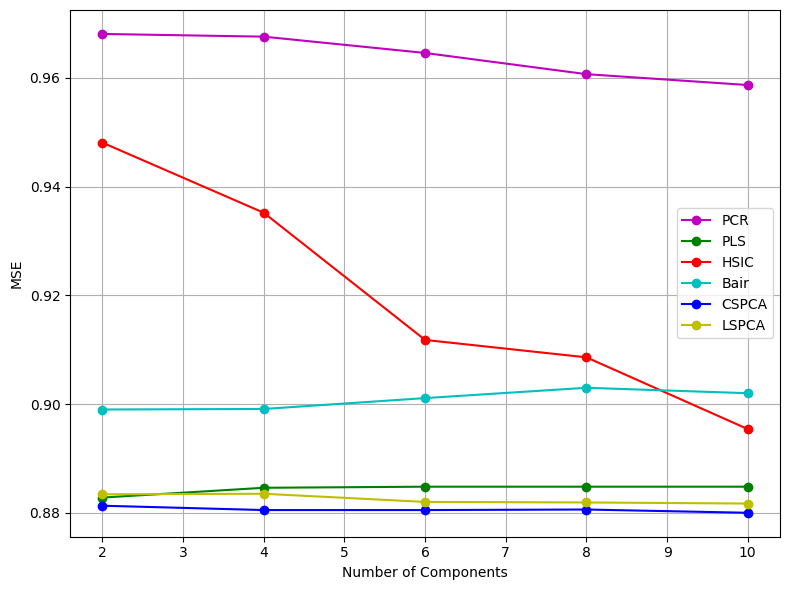}
        \caption{Mean Squared Error (Sim 2)}
        \label{Fig:2b}
    \end{subfigure}
    \hfill
    \begin{subfigure}{0.32\textwidth}
        \centering
        \includegraphics[width=\linewidth]{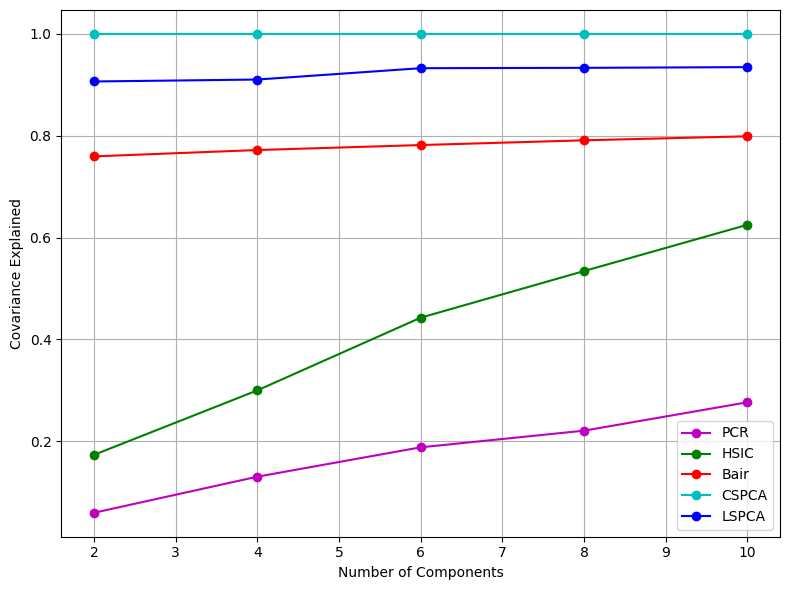}
        \caption{Covariance Explained (Sim 2)}
        \label{Fig:2c}
    \end{subfigure}

    \vspace{1em} % Add vertical spacing between rows

    % Row 3
    \begin{subfigure}{0.32\textwidth}
        \centering
        \includegraphics[width=\linewidth]{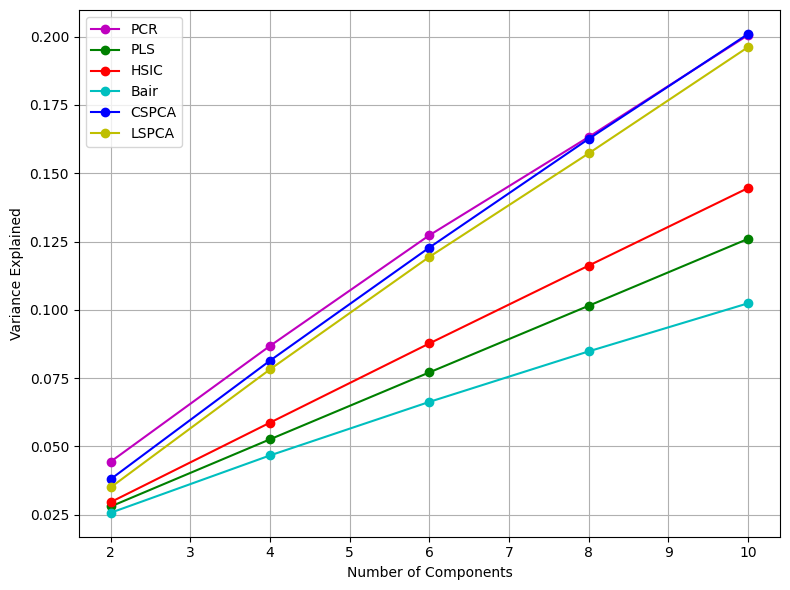}
        \caption{Variance Explained (Sim 3)}
        \label{Fig:3a}
    \end{subfigure}
    \hfill
    \begin{subfigure}{0.32\textwidth}
        \centering
        \includegraphics[width=\linewidth]{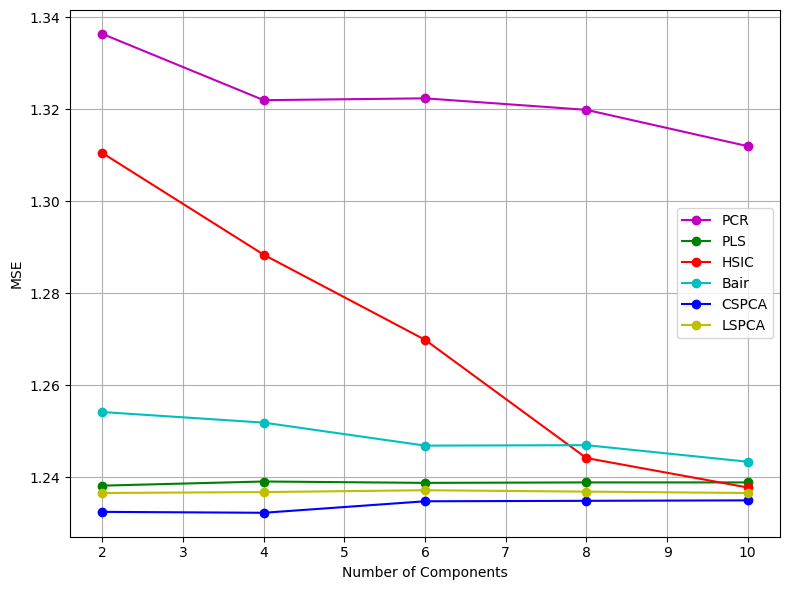}
        \caption{Mean Squared Error (Sim 3)}
        \label{Fig:3b}
    \end{subfigure}
    \hfill
    \begin{subfigure}{0.32\textwidth}
        \centering
        \includegraphics[width=\linewidth]{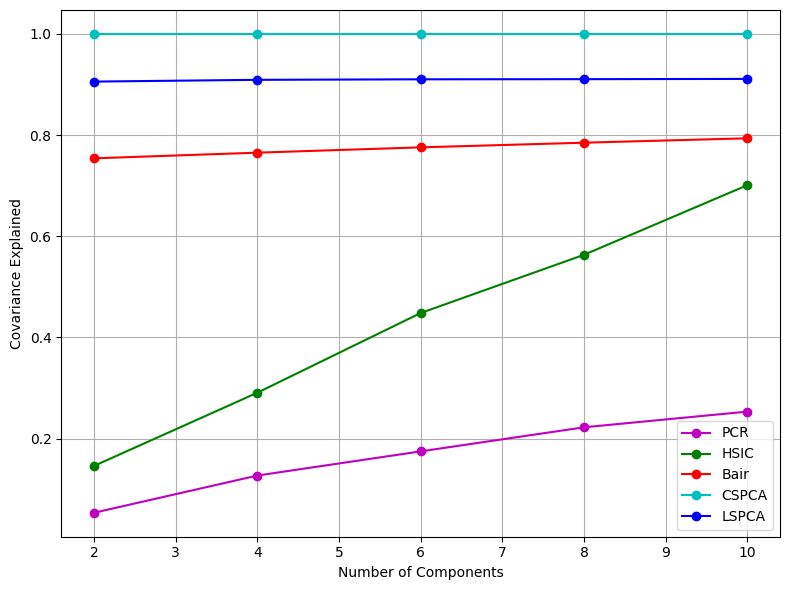}
        \caption{Covariance Explained (Sim 3)}
        \label{Fig:3c}
    \end{subfigure}

    \caption{Variance explained, Mean squared error, and Covariance explained for Simulation 1 (top), Simulation 2 (middle) and Simulation 3 (bottom) under the i.i.d. scenario.}
    \label{fig:iid}
\end{figure}

\begin{figure}[h]
    \centering
    % Row 1
    \begin{subfigure}{0.32\textwidth}
        \centering
        \includegraphics[width=\linewidth]{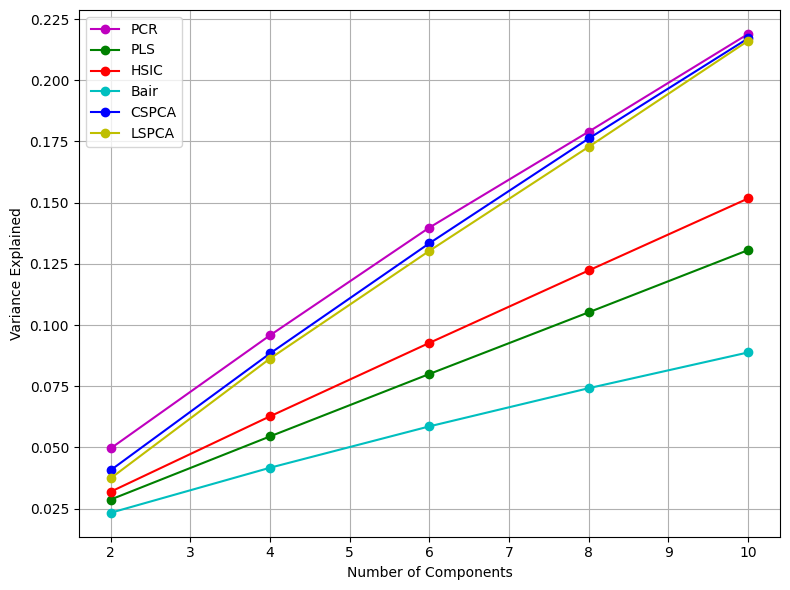}
        \caption{Variance Explained (Sim 1)}
        \label{Fig:1d}
    \end{subfigure}
    \hfill
    \begin{subfigure}{0.32\textwidth}
        \centering
        \includegraphics[width=\linewidth]{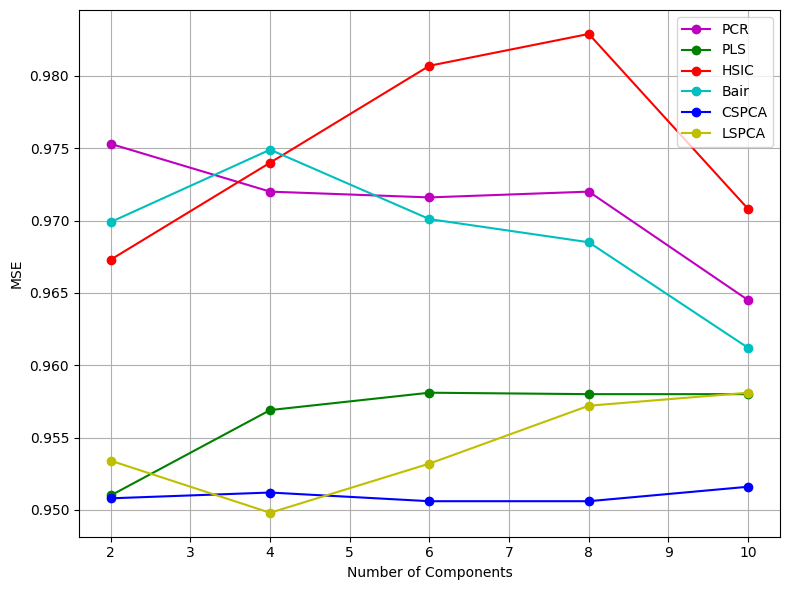}
        \caption{Mean Squared Error (Sim 1)}
        \label{Fig:1e}
    \end{subfigure}
    \hfill
    \begin{subfigure}{0.32\textwidth}
        \centering
        \includegraphics[width=\linewidth]{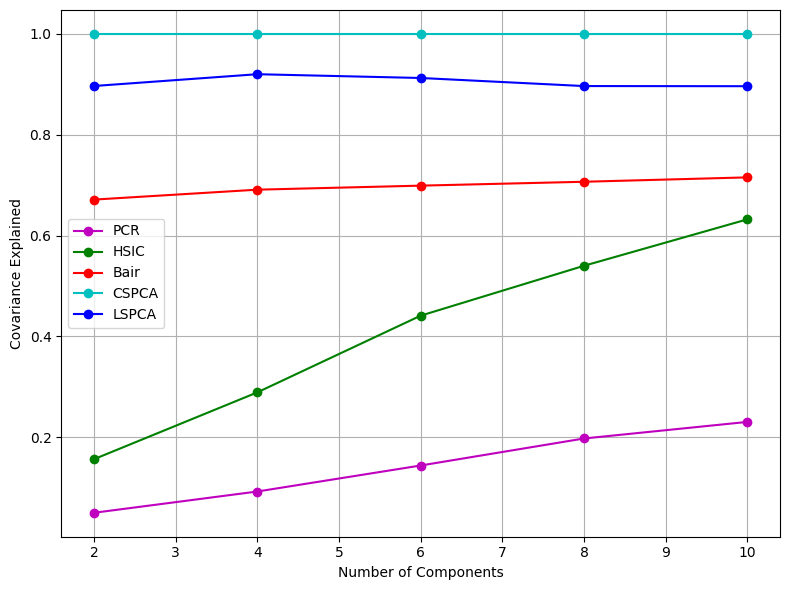}
        \caption{Covariance Explained (Sim 1)}
        \label{Fig:1f}
    \end{subfigure}

    \vspace{1em} % Add vertical spacing between rows

    % Row 2
    \begin{subfigure}{0.32\textwidth}
        \centering
        \includegraphics[width=\linewidth]{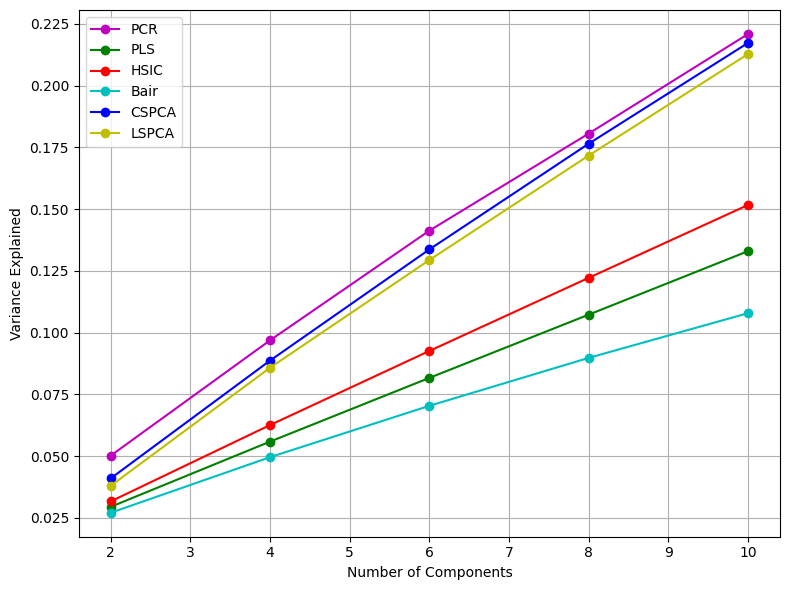}
        \caption{Variance Explained (Sim 2)}
        \label{Fig:2d}
    \end{subfigure}
    \hfill
    \begin{subfigure}{0.32\textwidth}
        \centering
        \includegraphics[width=\linewidth]{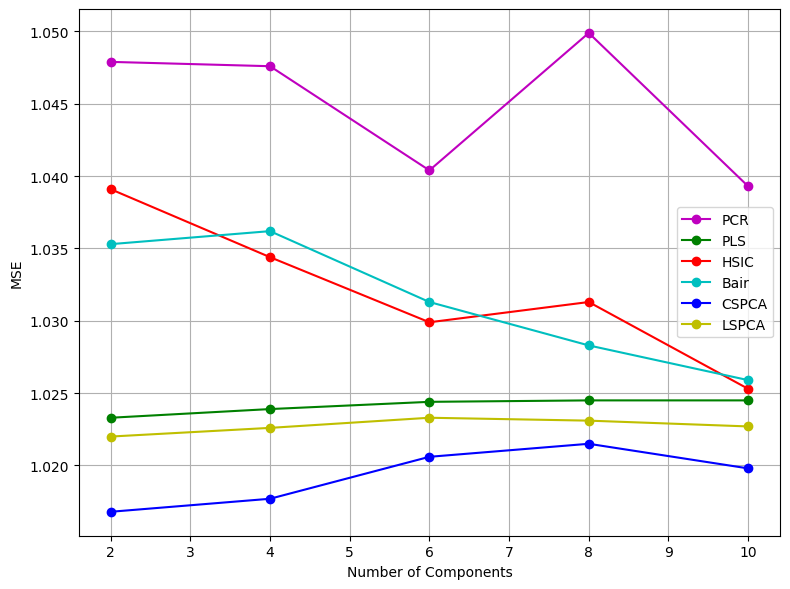}
        \caption{Mean Squared Error (Sim 2)}
        \label{Fig:2e}
    \end{subfigure}
    \hfill
    \begin{subfigure}{0.32\textwidth}
        \centering
        \includegraphics[width=\linewidth]{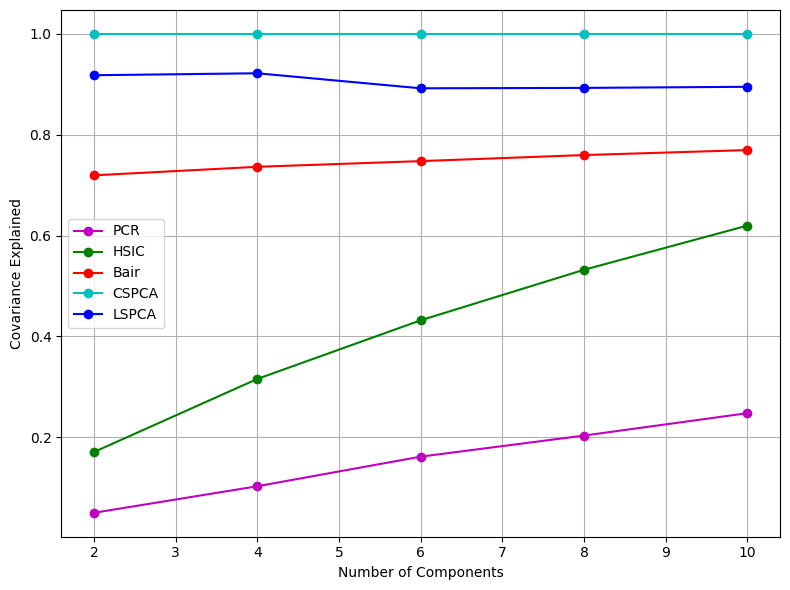}
        \caption{Covariance Explained (Sim 2)}
        \label{Fig:2f}
    \end{subfigure}

    \vspace{1em} % Add vertical spacing between rows

    % Row 3
    \begin{subfigure}{0.32\textwidth}
        \centering
        \includegraphics[width=\linewidth]{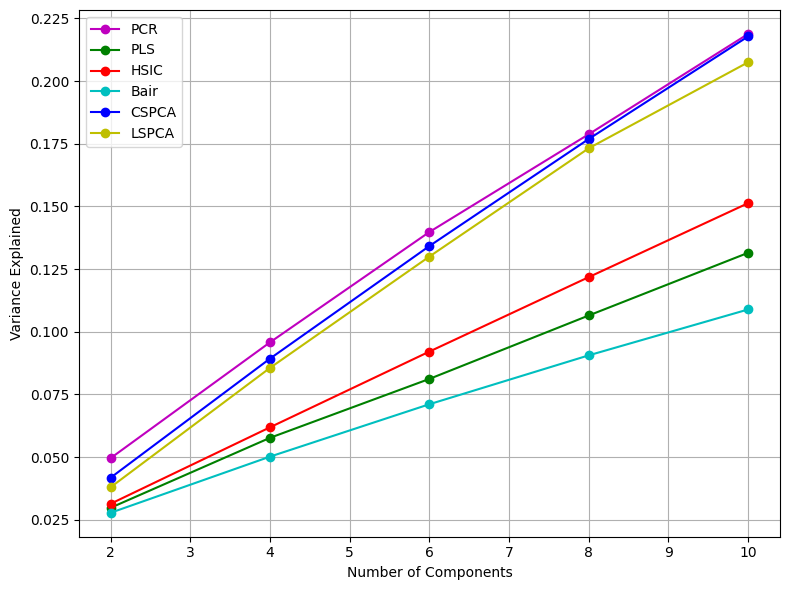}
        \caption{Variance Explained (Sim 3)}
        \label{Fig:3d}
    \end{subfigure}
    \hfill
    \begin{subfigure}{0.32\textwidth}
        \centering
        \includegraphics[width=\linewidth]{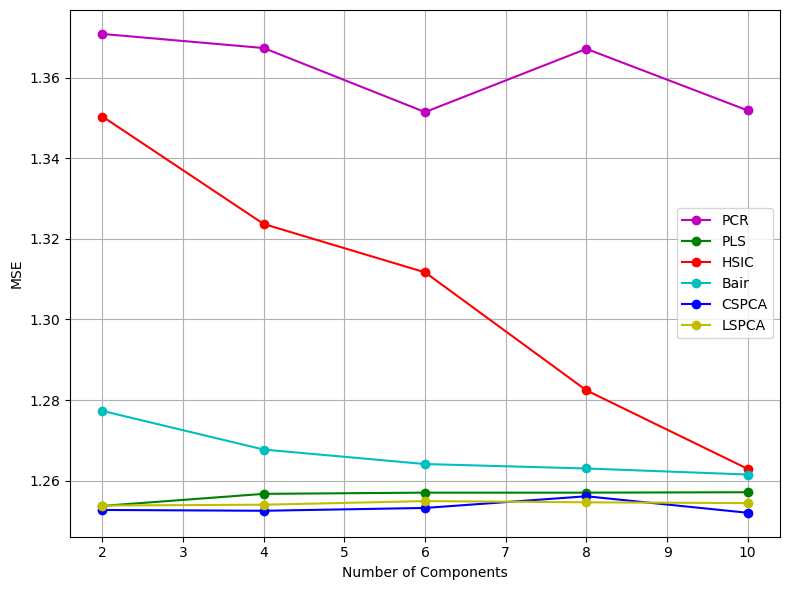}
        \caption{Mean Squared Error (Sim 3)}
        \label{Fig:3e}
    \end{subfigure}
    \hfill
    \begin{subfigure}{0.32\textwidth}
        \centering
        \includegraphics[width=\linewidth]{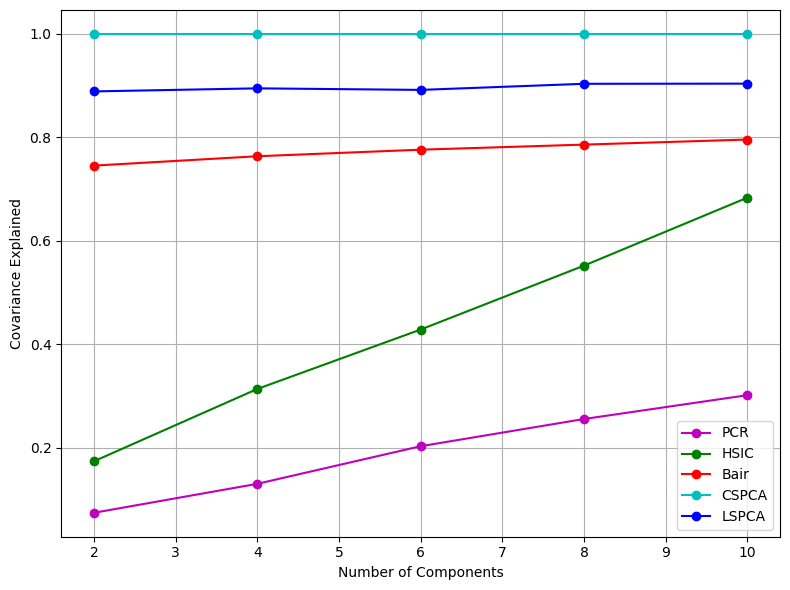}
        \caption{Covariance Explained (Sim 3)}
        \label{Fig:3f}
    \end{subfigure}

    \caption{Variance explained, Mean squared error, and Covariance explained for Simulation 1 (top), Simulation 2 (middle) and Simulation 3 (bottom) under the correlated scenario.}
    \label{fig:cor}
\end{figure}

In our final simulation, we investigate an even more complex relationship with the addition of a quadratic term. The i.i.d. scenario results are presented in Figure \ref{fig:iid} (bottom three plots), with corresponding correlated results shown in Figure \ref{fig:cor} (bottom three plots).

The findings in the i.i.d. setting align closely with the previous simulations. In terms of explained variance, CSPCA and LSPCA perform comparably to PCA, with CSPCA even outperforming PCA for $q=10$ components. For prediction error, CSPCA, PLS, and LSPCA remain competitive across all component numbers, while Bair’s method and HSIC-based SPCA show strong performance with larger component counts ($q=8,10$). CSPCA exhibits a slight but consistent advantage, achieving the lowest MSE across all component numbers. Regarding response relevance, CSPCA maintains the strongest performance, followed by LSPCA.

For the correlated scenario, conclusions for variance and covariance explained remain consistent with the i.i.d. scenario. However, the prediction error results differ: while CSPCA retains a small advantage over PLS and LSPCA across most component numbers, except $q=8$ components, the differences between methods are less pronounced.

\subsubsection{Simulation Discussion}
Overall, results across all simulation analyses, whether for the i.i.d. or correlated scenario, indicate similar patterns. Specifically, CSPCA and LSPCA are able to approach PCA's variance explained performance thanks to the regularisation parameter in their objective function, while still maintaining strong predictive performance. CSPCA and LSPCA indicate the strongest performance in terms of prediction error, along with PLS, with CSPCA indicating a small advantage for the majority of cases considered. Finally, CSPCA is able to preserve the covariance structure between the projected data and the response variable, providing relevant projections with respect to the response. LSPCA also performs well in terms of covariance explained, but its performance is slightly inferior. A notable observation is that supervised methods perform better than PCA in terms of prediction error and covariance explained since they incorporate the response variable in their framework, however they are inferior when it comes to variance explained. In conclusion, the addition of a balancing parameter in the objective function of the SPCA problem provides superior performance across all metrics compared to single-objective focused methods.

\subsection{Applications to real datasets}
We test the performance of CSPCA in four real world datasets, two for regression and two for classification tasks. For the former, we consider the low dimensional MUSIC dataset \citep{Zhou2014Predicting-the-} from the UCI Machine Learning Repository\footnote{\url{https://archive.ics.uci.edu/dataset/315/geographical+original+of+music}} that contains audio features of $n=1058$ songs, consists of two response variables and has been used in the past to compare SPCA methods \citep{Ritchie2019Supervised-Prin}, along with a high dimensional dataset that originates from a microarray experiment for liver toxicity where $n=64$ rats were exposed to different doses of acetaminophen \citep{Bushel2007Simultaneous-cl}. The latter was extracted from the mixOmics library from the Bioconductor R Package \footnote{\url{https://www.rdocumentation.org/packages/mixOmics/versions/6.3.2/topics/liver.toxicity}}.

The classification datasets we consider are also high dimensional and originate from DNA microarray experiments. The first one is the leukemia dataset by \citet{Golub1999Molecular-class}, which is used to classify $n=72$ patients with two types of leukemia. The second is a colon cancer dataset from \citet{Alon1999Broad-patterns-}, which is used to classify two types of cancer tissues for $n=62$ individuals. The performance metrics we consider for these two datasets are variance explained, logistic loss, accuracy and area under the curve (AUC).

\subsubsection{The MUSIC Dataset}\label{MUSIC}
The MUSIC dataset, originally introduced by \citet{Zhou2014Predicting-the-}, comprises 1,058 musical observations characterized by 116 audio features, along with two response variables encoding geographical origin information. We evaluated performance of SPCA methods, along with PLS, through 20 independent random splits ($80\%$ training, $20\%$ test), reporting Monte Carlo averages of the results across all splits. All methods except Bair's (inapplicable for multiple responses) were applied across a range of component numbers, with detailed numerical results presented in Table \ref{tab:resultsMUSIC}. 

CSPCA demonstrates superior performance in covariance explained across all component numbers, while also outperforming supervised methods in variance explained, approaching PCA's performance levels. For prediction accuracy, CSPCA achieves lower MSE than both PCR and SPCA using HSIC universally, and outperforms all methods for $q=2$ and $10$ components, while also surpassing LSPCA for $q=4$ components. CSPCA maintains an optimal balance across evaluation metrics, preserving PCA's interpretability while approaching (or surpassing) PLS's predictive power. Most significantly, CSPCA projections prove to be the most response-informative, as evidenced by its consistently superior covariance explained performance.

\begin{table*}[t]
\centering
\normalsize % Larger font to utilize space
\setlength{\tabcolsep}{5pt}
\resizebox{\textwidth}{!}{%
\begin{tabular}{l*{5}{S[table-format=1.4]@{\,±\,}S[table-format=1.4]}}
\toprule
\multirow{2}{*}{Methods} & \multicolumn{10}{c}{Number of Components} \\ % Adjusted header
\cmidrule(lr){2-11}
 & \multicolumn{2}{c}{2} & \multicolumn{2}{c}{4} & \multicolumn{2}{c}{6} & \multicolumn{2}{c}{8} & \multicolumn{2}{c}{10} \\
\midrule
\multicolumn{11}{l}{\textbf{Variance Explained}} \\
PCR & 0.4901 & 0.0006 & 0.6114 & 0.0005 & 0.6847 & 0.0004 & 0.7299 & 0.0004 & 0.7640 & 0.0004 \\
PLS & 0.3613 & 0.0018 & 0.4760 & 0.0013 & 0.5179 & 0.0016 & 0.5404 & 0.0013 & 0.5655 & 0.0020 \\
HSIC & 0.4584 & 0.0037 & 0.5787 & 0.0007 & 0.6383 & 0.0006 & 0.6876 & 0.0010 & 0.7224 & 0.0008 \\
CSPCA & 0.4641 & 0.0019 & 0.5949 & 0.0006 & 0.6722 & 0.0004 & 0.7187 & 0.0004 & 0.7558 & 0.0003 \\
LSPCA & 0.4732 & 0.0085 & 0.5675 & 0.0077 & 0.6400 & 0.0062 & 0.6892 & 0.0019 & 0.7300 & 0.0004 \\
\midrule
\multicolumn{11}{l}{\textbf{MSE}} \\
PCR & 0.9749 & 0.0242 & 0.8567 & 0.0163 & 0.8526 & 0.0209 & 0.8662 & 0.0208 & 0.8702 & 0.0243 \\
PLS & 0.8984 & 0.0227 & 0.8225 & 0.0149 & 0.7875 & 0.0182 & 0.8124 & 0.0194 & 0.8059 & 0.0222 \\
HSIC & 0.9628 & 0.0245 & 0.8581 & 0.0168 & 0.8482 & 0.0207 & 0.8525 & 0.0207 & 0.8512 & 0.0241 \\
CSPCA & 0.8880 & 0.0226 & 0.8372 & 0.0159 & 0.8328 & 0.0198 & 0.8427 & 0.0194 & 0.8366 & 0.0220 \\
LSPCA & 0.9632 & 0.0250 & 0.8496 & 0.0177 & 0.8044 & 0.0205 & 0.8182 & 0.0217 & 0.8115 & 0.0241 \\
\midrule
\multicolumn{11}{l}{\textbf{Covariance Explained}} \\
PCR & 0.6592 & 0.0043 & 0.9349 & 0.0008 & 0.9474 & 0.0014 & 0.9485 & 0.0008 & 0.9562 & 0.0009 \\
%PLS & {-} & {-} & {-} & {-} & {-} & {-} & {-} & {-} & {-} & {-} \\
HSIC & 0.7191 & 0.0131 & 0.9355 & 0.0016 & 0.9593 & 0.0017 & 0.9793 & 0.0007 & 0.9867 & 0.0003 \\
CSPCA &1.0000 & 0.0000& 1.0000 & 0.0000 & 1.0000 & 0.0000 & 1.0000 & 0.0000 & 1.0000 & 0.0000 \\
LSPCA & 0.6433 & 0.0162 & 0.9071 & 0.0186 & 0.9507 & 0.0023 & 0.9533& 0.0020 & 0.9538 & 0.0010 \\
\bottomrule
\end{tabular}
}
\caption{Monte Carlo estimates of three metrics ± SE (standard error) across varying numbers of components for the MUSIC dataset with 20 random splits into 80\%--20\% training and test sets.}
\label{tab:resultsMUSIC}
\end{table*}

\subsubsection{The MICE Genetic Dataset}\label{RNA}
The liver toxicity dataset, first presented by \citet{Bushel2007Simultaneous-cl}, comprises gene expression profiles of $p=3116$ genes measured in $n=64$ male rats exposed to varying doses of acetaminophen (paracetamol), along with 10 clinical response variables. We focus on Albumin levels (ALB.g.dl) as our response variable. Given the high dimensionality ($p = 3116 \gg n = 64$), we implement CSPCA using Nyström's approximation with $m = 55$ ($\sqrt{3116} \approx 55.82$). We evaluated performance of SPCA methods, along with PLS, through 20 independent random splits ($80\%$ training, $20\%$ test), reporting Monte Carlo averages of the results across all splits. All methods were applied across a range of component numbers, with detailed numerical results presented in Table \ref{tab:resultsMICE}).

CSPCA demonstrates consistent superiority in covariance explained across all component numbers while approaching, along with LSPCA, PCA's variance explained performance, both significantly outperforming other methods. SPCA using HSIC and Bair's method illustrate strong performance in terms of covariance explained as the number of components increases, surpassing both LSPCA and PCA. For prediction error, CSPCA shows robust MSE performance at all dimensionality levels, outperforming all SPCA variants for $q=2,4,6$ and $8$ components and PLS for $q=4$ components. These results confirm CSPCA's ability to maintain balanced performance across all evaluation metrics without trade-offs between variance explained, prediction accuracy, or response relevance.

\begin{table*}[t]
\centering
\normalsize % Larger font to utilize space
\setlength{\tabcolsep}{4pt}
\resizebox{\textwidth}{!}{%
\begin{tabular}{l*{5}{S[table-format=1.4]@{\,±\,}S[table-format=1.4]}}
\toprule
\multirow{2}{*}{Methods} & \multicolumn{10}{c}{Number of Components} \\ % Adjusted header
\cmidrule(lr){2-11}
 & \multicolumn{2}{c}{2} & \multicolumn{2}{c}{4} & \multicolumn{2}{c}{6} & \multicolumn{2}{c}{8} & \multicolumn{2}{c}{10} \\
\midrule
\multicolumn{11}{l}{\textbf{Variance Explained}} \\
PCR & 0.4340 & 0.0017& 0.5705 & 0.0014& 0.6616 & 0.0008& 0.7232 & 0.0007& 0.7671 & 0.0007\\
PLS & 0.2128 & 0.0053& 0.3418 & 0.0042& 0.3984 & 0.0023& 0.4343 & 0.0023& 0.4710 & 0.0024\\
HSIC & 0.3575 & 0.0037& 0.4706 & 0.0023& 0.5349 & 0.0025& 0.5802 & 0.0023& 0.6222 & 0.0026 \\
Bair & 0.2882 & 0.0025& 0.3910 & 0.0020& 0.4525 & 0.0023& 0.4911 & 0.0023& 0.5222 & 0.0028\\
CSPCA & 0.4071 & 0.0026& 0.5693 & 0.0014& 0.6599 & 0.0009& 0.7216 & 0.0007& 0.7654 & 0.0007 \\
LSPCA & 0.4271 & 0.0025& 0.5607 & 0.0022& 0.6486 & 0.0022& 0.7120 & 0.0035& 0.7534 & 0.0030
 \\
\midrule
\multicolumn{11}{l}{\textbf{MSE}} \\
PCR &1.0727 & 0.0617& 0.6196 & 0.0465 &0.6372 & 0.0385& 0.6423 & 0.0396& 0.7442 & 0.0488\\
PLS & 0.6832 & 0.0396& 0.7418 & 0.0555& 0.6072 & 0.0360& 0.5766 & 0.0358& 0.6272 & 0.0452 \\
HSIC & 0.9876 & 0.0622& 0.6140 & 0.0466& 0.6376 & 0.0440& 0.6521 & 0.0385& 0.6769 & 0.0431 \\
Bair & 0.9271 & 0.0615& 0.6095 & 0.0457& 0.6419 & 0.0375& 0.6384 & 0.0371& 0.7413 & 0.0484 \\
CSPCA & 0.7010 & 0.0473& 0.5991 & 0.0430& 0.6339 & 0.0371& 0.6312 & 0.0344& 0.7355 & 0.0012 \\
LSPCA & 0.8299 & 0.1533& 0.6195 & 0.0464& 0.6643 & 0.0596& 0.7226 & 0.0948& 0.7576 & 0.0844
 \\
\midrule
\multicolumn{11}{l}{\textbf{Covariance Explained}} \\
PCR & 0.2680 & 0.0183& 0.8575 & 0.0064& 0.9125 & 0.0036& 0.9335 & 0.0020& 0.9448 & 0.0020\\
%PLS & {-} & {-} & {-} & {-} & {-} & {-} & {-} & {-} & {-} & {-} \\
HSIC & 0.5359 & 0.0264 & 0.8696 & 0.0070& 0.9555 & 0.0035& 0.9789 & 0.0016& 0.9925 & 0.0009\\
Bair & 0.5343 & 0.0157& 0.9015 & 0.0035& 0.9304 & 0.0020& 0.9409 & 0.0015& 0.9472 & 0.0016 \\
CSPCA & 0.9725 & 0.0042 & 0.9919 & 0.0010 & 0.9946 & 0.0005 & 0.9960 & 0.0003 &0.9972 & 0.0003\\
LSPCA & 0.2821 & 0.0237& 0.8939 & 0.0089& 0.8987 & 0.0133& 0.9382 & 0.0022& 0.9438 & 0.0039
\\
\bottomrule
\end{tabular}
}
\caption{Monte Carlo estimates of three metrics ± SE (standard error) across varying numbers of components for the mice liver toxicity dataset by Bushel et al. with 20 splits into 80\%--20\% training and test sets.}
\label{tab:resultsMICE}
\end{table*}

\subsubsection{The Leukemia Genetic Dataset}
The leukemia genetic dataset originates from \citet{Golub1999Molecular-class} and is publicly available on Kaggle \footnote{\url{https://www.kaggle.com/datasets/crawford/gene-expression}}. It contains $p=7129$ gene expression levels for $n=72$ individuals. The dataset is used to classify patients into two types of leukemia, acute myeloid leukemia (AML) and acute lymphoblastic leukemia (ALL). The dataset is already divided into training ($n_{\text{tr}}=38$) and test ($n_{\text{te}}=34$) sets and hence comparison is applied directly on this division. Again, we use the extension of CSPCA using Nyström's method to enhance scalability. Numerical results from our analysis are presented in Table \ref{tab:resultsCL}. CSCPA outperforms all methods in terms of logistic loss and AUC across all number of components. SPCA using HSIC also shows strong but inferior performance compared to CSPCA in terms of logistic loss and AUC, along with Bair's method for some components. LSPCA does not perform so well in terms of logistic loss for the lower components but improves as the number of components increases. PCA and LDA are the worst performing methods in terms of both logistic loss and AUC. In terms of accuracy, CSPCA performs consistently well for all components and is always the best or tied-best performing method. SPCA using HSIC, along with Bair's method, offer competitive results as well, while LSPCA also performs well except for $q=2$ components. Finally, in terms of variance explained, CSPCA along with LSPCA approaches PCA's performance, with other methods being substantially inferior. In conclusion, CSPCA offers a great balance across all performance metrics. It provides the best performance in terms of logistic loss, AUC and accuracy for most components, while also preserving a significant amount of variance, similar to PCA.

\begin{table}[h]
\centering
\small
\setlength{\tabcolsep}{4pt}
\begin{tabular}{l*{5}{S[table-format=1.4]}}
\toprule
\multirow{2}{*}{Methods} & \multicolumn{5}{c}{Number of Components} \\
\cmidrule(lr){2-6}
 & {2} & {4} & {6} & {8} & {10} \\
\midrule
\multicolumn{6}{l}{\textbf{Variance Explained}} \\
PCR   & 0.2697 & 0.3845 & 0.4680 & 0.5358 & 0.5919 \\
%LDA   & {-}    & {-}    & {-}    & {-}    & {-}    \\
HSIC  & 0.0904 & 0.0907 & 0.0909 & 0.0913 & 0.0915 \\
Bair  & 0.0534 & 0.0685 & 0.0771 & 0.0841 & 0.0895 \\
CSPCA & 0.2618 & 0.3741 & 0.4471 & 0.5107 & 0.5574 \\
LSPCA & 0.2597 & 0.3664 & 0.4553 & 0.5122 & 0.5317 \\
\midrule
\multicolumn{6}{l}{\textbf{Logistic Loss}} \\
PCR   & 0.6209 & 0.6888 & 0.6560 & 0.7908 & 0.5347 \\
LDA   & 0.5346 & 0.5346 & 0.5346 & 0.5346 & 0.5346 \\
HSIC  & 0.4332 & 0.3212 & 0.2490 & 0.3918 & 0.3870 \\
Bair  & 0.5300 & 0.3302 & 0.2612 & 0.3844 & 0.3847 \\
CSPCA & 0.4052 & 0.2930 & 0.2484 & 0.3302 & 0.3723 \\
LSPCA & 0.6115 & 0.5275 & 0.4276 & 0.3720 & 0.3809 \\
\midrule
\multicolumn{6}{l}{\textbf{Accuracy}} \\
PCR   & 0.6176 & 0.8529 & 0.8529 & 0.7647 & 0.8235 \\
LDA   & 0.7059 & 0.7059 & 0.7059 & 0.7059 & 0.7059 \\
HSIC  & 0.8235 & 0.8529 & 0.8529 & 0.8529 & 0.8235 \\
Bair  & 0.8529 & 0.8824 & 0.8529 & 0.7941 & 0.7941 \\
CSPCA & 0.8529 & 0.8824 & 0.8824 & 0.8529 & 0.8235 \\
LSPCA & 0.7059 & 0.8824 & 0.8529 & 0.7941 & 0.8235 \\
\midrule
\multicolumn{6}{l}{\textbf{AUC}} \\
PCR   & 0.7821 & 0.9429 & 0.9429 & 0.9464 & 0.9571 \\
LDA   & 0.8964 & 0.8964 & 0.8964 & 0.8964 & 0.8964 \\
HSIC  & 0.9500 & 0.9429 & 0.9321 & 0.9429 & 0.9464 \\
Bair  & 0.9286 & 0.9607 & 0.9679 & 0.9714 & 0.9857 \\
CSPCA & 0.9857 & 0.9821 & 0.9821 & 0.9964 & 0.9893 \\
LSPCA & 0.8107 & 0.9536 & 0.9607 & 0.8250 & 0.9643 \\
\bottomrule
\end{tabular}
\caption{Results across varying numbers of components for the leukemia dataset by Golub et al.. The data were already split into a training and a test set.}
\label{tab:resultsCL}
\end{table}

\subsubsection{The Colon Cancer Dataset}
The colon cancer dataset originates from \citet{Alon1999Broad-patterns-} and consists of $p=2000$ gene expression levels from $n=62$ individuals. It was extracted from the colonCA library from the Bioconductor R package \footnote{\url{https://bioconductor.org/packages/release/data/experiment/html/colonCA.html}}. The dataset consists of two classes corresponding to tissue identity, normal and tumor tissue. An 80\%--20\% split into training and test sets was applied 20 different times and results across all splits were averaged. Numerical results from our analysis are presented in Table \ref{tab:resultsCANC}. CSPCA outperforms all methods in terms of logistic loss across almost all number of components, except $q=8$ components where LSPCA outperforms CSPCA, provides the best accuracy results for $q=6,8$ and $10$ components and second best results for $q=2$ and $4$ components, showing exceptional prediction accuracy. CSPCA also provides the best results in terms of AUC for $q=2,4,6$ and $10$ components and second best for $q=8$ components, while Bair's method provides the highest AUC score for $q=8$ components. Finally, in terms of variance explained, CSPCA, along with LSPCA are approaching the performance of PCA, with remaining methods being significantly inferior. Overall, similar to the leukemia dataset, CSPCA offers the best balance of preserving most of the variance of the original data while also providing top-performing results in terms of classification.

\begin{table*}[t]
\centering
\normalsize % Larger font to utilize space
\setlength{\tabcolsep}{4pt}
\resizebox{\textwidth}{!}{%
\begin{tabular}{l*{5}{S[table-format=1.4]@{\,±\,}S[table-format=1.4]}}
\toprule
\multirow{2}{*}{Methods} & \multicolumn{10}{c}{Number of Components} \\ % Adjusted header
\cmidrule(lr){2-11}
 & \multicolumn{2}{c}{2} & \multicolumn{2}{c}{4} & \multicolumn{2}{c}{6} & \multicolumn{2}{c}{8} & \multicolumn{2}{c}{10} \\
\midrule
\multicolumn{11}{l}{\textbf{Variance Explained}} \\
PCR & 0.5507 & 0.0045& 0.6782 & 0.0037& 0.7464 & 0.0027& 0.7947 & 0.0022& 0.8286 & 0.0019\\
%LDA & {-} & {-} & {-} & {-} & {-} & {-} & {-} & {-} & {-} & {-} \\
HSIC & 0.1495 & 0.0132 & 0.1503 & 0.0131 & 0.1510 & 0.0131 & 0.1518 & 0.0131 & 0.1525 & 0.0131  \\
Bair & 0.1930 & 0.0080 & 0.2398 & 0.0094 & 0.2614 & 0.0101 & 0.2766 & 0.0105 & 0.2879 & 0.0109  \\
CSPCA & 0.5438 & 0.0053& 0.6702 & 0.0040& 0.7341 & 0.0032& 0.7803 & 0.0026& 0.8111 & 0.0023\\
LSPCA &  0.5410 & 0.0074 & 0.6579 & 0.0073 & 0.7337 & 0.0045 & 0.7789 & 0.0047 & 0.8110 & 0.0028\\
\midrule
\multicolumn{11}{l}{\textbf{Logistic Loss}} \\
PCR &0.7106 & 0.0183 &0.5944 & 0.0586& 0.6409 & 0.0760& 0.6893 & 0.0966& 0.8588 & 0.1585\\
LDA & 0.9682 & 0.0985& 0.9682 & 0.0985& 0.9682 & 0.0985& 0.9682 & 0.0985&0.9682 & 0.0985 \\
HSIC & 0.5690 & 0.0616& 0.6102 & 0.0835& 0.7294 & 0.1048& 0.9236 & 0.1735& 1.1833 & 0.2152 \\
Bair & 0.6161 & 0.0534& 0.5873 & 0.0637& 0.6220 & 0.0812& 0.7025 & 0.1140& 0.8363 & 0.1417 \\
CSPCA & 0.5624 & 0.0614 & 0.5865 & 0.0725 & 0.5788 & 0.0709 & 0.6574 & 0.0819 & 0.8000 & 0.1404 \\
LSPCA &  0.6697 & 0.0118 & 0.5944 & 0.0582 & 0.6138 & 0.0675 & 0.6359 & 0.0776 & 0.8450 & 0.1355\\
\midrule
\multicolumn{11}{l}{\textbf{Accuracy}} \\
PCR & 0.5577 & 0.0162& 0.7346 & 0.0240& 0.7731 & 0.0234 & 0.7962 & 0.0257& 0.7923 & 0.0273\\
LDA & 0.7654 & 0.0158 & 0.7654 & 0.0158 & 0.7654 & 0.0158 & 0.7654 & 0.0158 & 0.7654 & 0.0158 \\
HSIC & 0.8231 & 0.0250& 0.8192 & 0.0257& 0.8115 & 0.0214& 0.7962 & 0.0226& 0.8000 & 0.0257\\
Bair & 0.7038 & 0.0354&0.8077 & 0.0269& 0.7923 & 0.0218& 0.8038 & 0.0214& 0.7962 & 0.0273\\
CSPCA & 0.8192 & 0.0239 & 0.8154 & 0.0252& 0.8231 & 0.0238& 0.8192 & 0.0226& 0.8077 & 0.0258 \\
LSPCA & 0.6000 & 0.0215 & 0.7385 & 0.0233 & 0.7538 & 0.0270 & 0.8115 & 0.0264 & 0.7731 & 0.0234
\\
\midrule
\multicolumn{11}{l}{\textbf{AUC}} \\
PCR & 0.4388 & 0.0372& 0.8187 & 0.0277& 0.8175 & 0.0268& 0.8475 & 0.0238& 0.8387 & 0.0286\\
LDA &  0.7937 & 0.0200 & 0.7937 & 0.0200 & 0.7937 & 0.0200 & 0.7937 & 0.0200 & 0.7937 & 0.0200 \\
HSIC & 0.8563 & 0.0231& 0.8192 & 0.0257& 0.8115 & 0.0214& 0.8475 & 0.0236& 0.8337 & 0.0255\\
Bair & 0.7162 & 0.0532& 0.8400 & 0.0269& 0.8500 & 0.0226& 0.8638 & 0.0192& 0.8562 & 0.0235\\
CSPCA & 0.8600 & 0.0216& 0.8562 & 0.0249 & 0.8550 & 0.0233& 0.8525 & 0.0238 & 0.8575 & 0.0245 \\
LSPCA & 0.6162 & 0.0360 & 0.8238 & 0.0212 & 0.8175 & 0.0279 & 0.8250 & 0.0273 & 0.8225 & 0.0296
\\
\bottomrule
\end{tabular}
}
\caption{Monte Carlo estimates of three metrics ± SE (standard error)  across varying numbers of components for the colon cancer dataset by Alon et al. with 20 splits into 80\%--20\% training and test sets.}
\label{tab:resultsCANC}
\end{table*}

\subsection{Data Visualization}
We now test the performance of CSPCA compared to existing SPCA methods in data visualization tasks. We begin with the simple example of the Iris dataset \footnote{\url{https://archive.ics.uci.edu/dataset/53/iris}} and then we use the leukemia dataset by \citet{Golub1999Molecular-class}. A 70\%--30\% split into training and test sets was performed in both datasets followed by scaling of the feature data.

\begin{figure}
    \centering
    \includegraphics[width=1\linewidth]{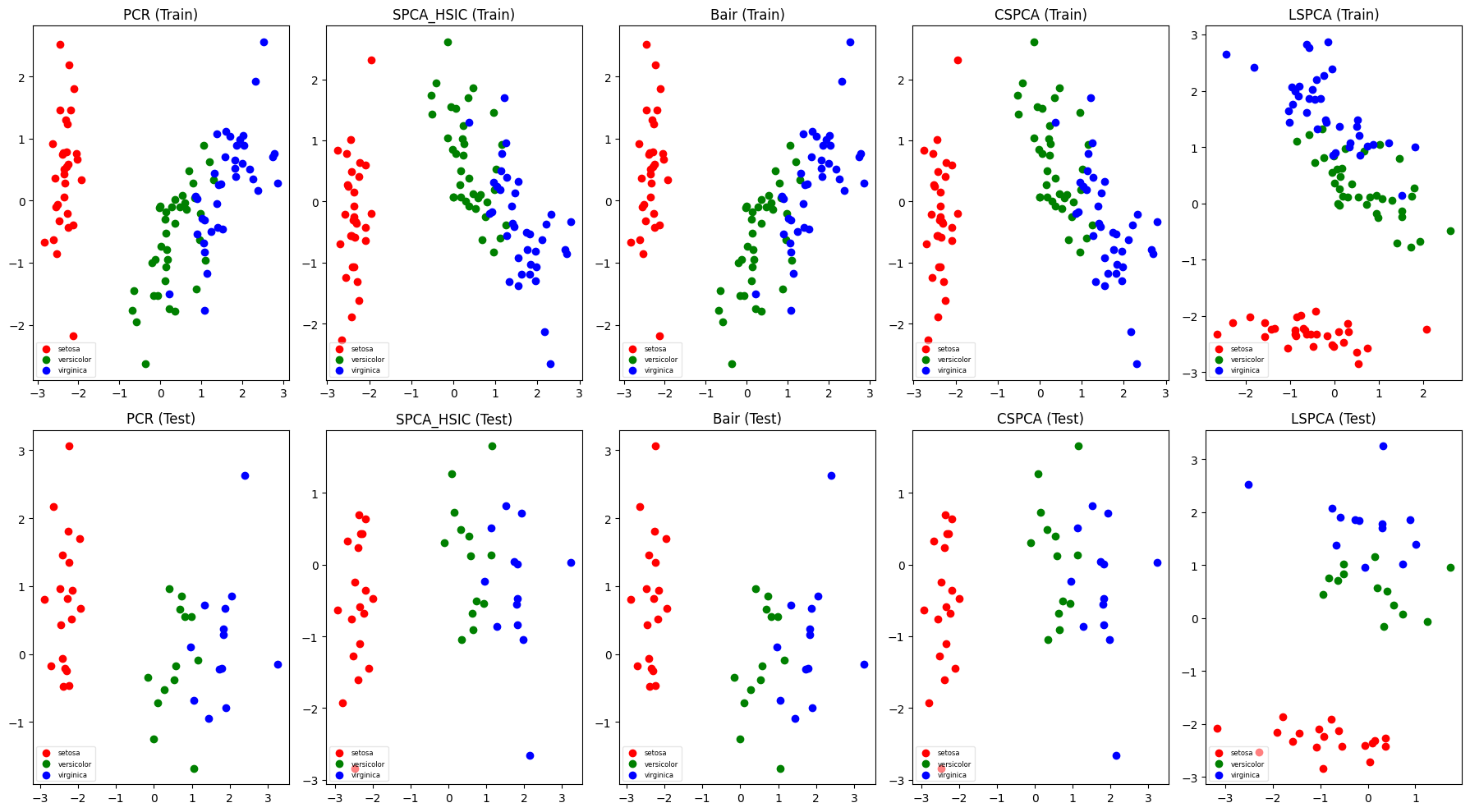}
    \caption{Two dimensional projection of the Iris dataset produced by PCR, SPCA using HSIC, Bair's method, CSPCA and LSPCA.}
    \label{fig:Iris}
\end{figure}

Figure \ref{fig:Iris} illustrates the results for the Iris dataset. The figures on top illustrate the projections into a two dimensional space for the training data and the figures below for the test data. None of the methods achieve excellent separability as there is consistent overlap between versicolor and virginica. However, setosa is consistently the most separated class in all methods. For the training set, all methods perform equal except PCR that shows a higher overlap compared to other methods. For the test set, there is still overlap between versicolor and virginica samples, although this is less apparent in SPCA using HSIC and CSPCA. Overall, we see that CSPCA works equally well or outperforms exisitng methods in terms of separability in the baseline example of the iris dataset.

\begin{figure}
    \centering
    \includegraphics[width=1\linewidth]{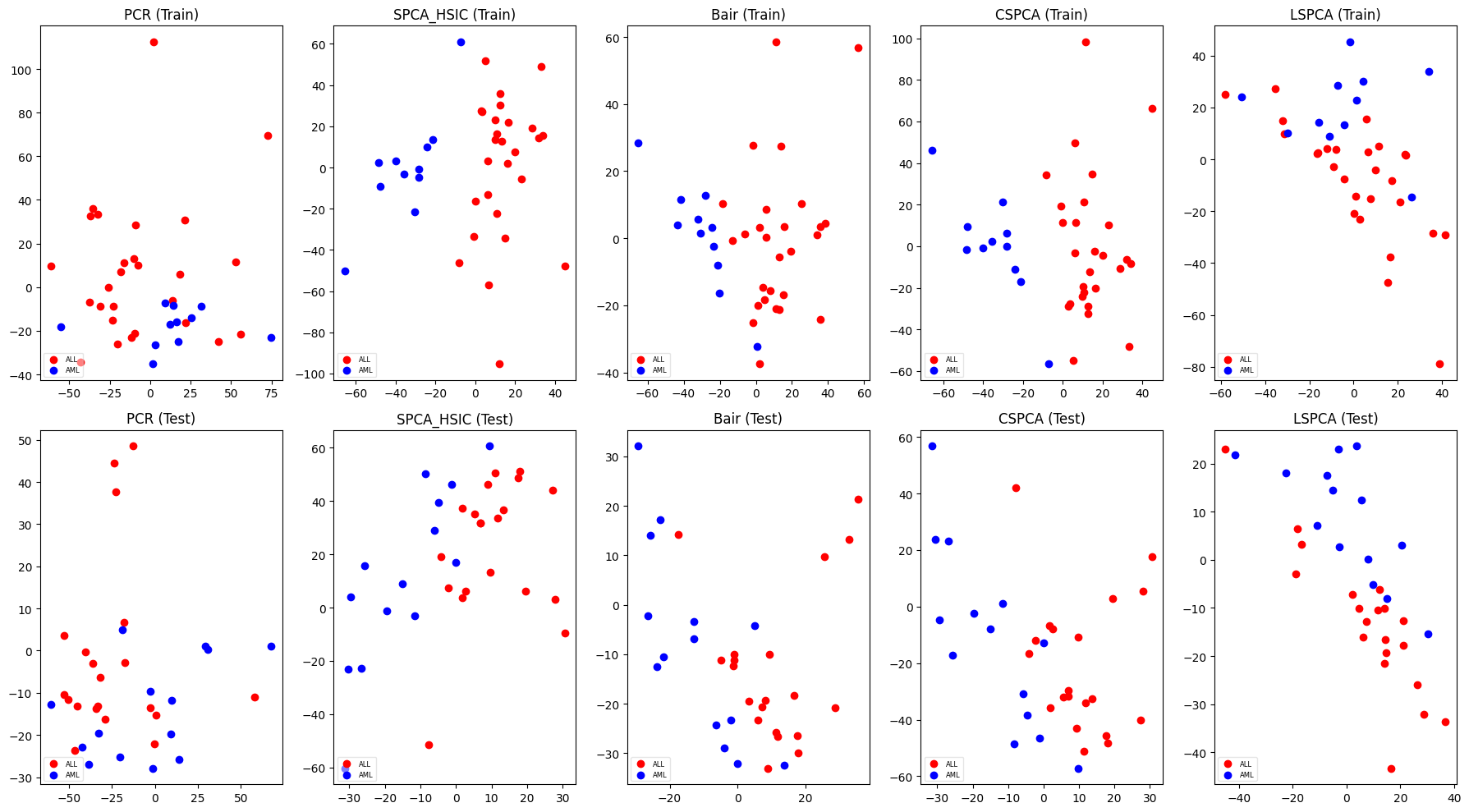}
    \caption{Two dimensional projection of Golub's leukemia dataset produced by PCR, SPCA using HSIC, Bair's method, CSPCA and LSPCA.}
    \label{fig:Golub}
\end{figure}

Figure \ref{fig:Golub} illustrates the results for the leukemia dataset by \citet{Golub1999Molecular-class}. The blue points correspond to AML leukemia and the red points to ALL leukemia. Regarding the training data, we observe that SPCA using HSIC and CSPCA provide a clear separation between AML and ALL samples. Bair's method also performs well, although there appears an AML sample mixed with some ALL samples. PCR and LSPCA do not separate well the data as there is clear overlap between AML and ALL samples. Regarding the test data, we observe that once again SPCA using HSIC and CSPCA offer the best separation between the two leukemia types with one overlapping sample. Bair's method along with LSPCA also try to separate the data however the amount of overlap is clearly larger that the former methods. Finally, PCA provides the worst separation between the test data as there is significant overlap between the two leukemia types.

Overall, these two examples establish CSPCA's strong and competitive performance in visualization tasks as well against the state-of-the-art SPCA methods. Especially the performance in the real leukemia dataset provides promising applications of CSPCA in visualization tasks in domain-specific high dimensional genetic datasets.

\section{Discussion}
In this paper, we propose CSPCA, a novel supervised PCA dimensionality reduction technique that seeks to balance relevance and interpretability by simultaneously maximising the covariance between the response variables and the projected data and the variance explained by the projected data. As far as we are concerned, this is the first time these two competing objectives are injected into a single optimization framework, balanced by a tunable hyperparameter $\kappa$, while also providing a closed-form solution via eigenvalue decomposition. We extend the proposed method for cases where the number of features is very large---thousands to tens of thousands, by using Nyström's method for low-rank approximation of the objective matrix, effectively enhancing scalability while also preserving interpretability and relevance.

We have established CSPCA's practical performance in a number of simulation analyses as well as four real-world datasets for both regression and classification tasks. Notably, three of these datasets, which are high-dimensional with a large number of features, are domain-specific to statistical genetics, derived from microarray experiments. We have illustrated CSPCA's superior or equal performance against state-of-the-art SPCA methods and well-known baselines for a number of performance metrics. We have also presented applications of CSPCA for data visualization tasks, where CSPCA is able to compete or outperform existing methods in terms of separability of projected data from different classes.

CSPCA is based on a mathematically simple but rigorous framework, it is straightforward to implement while also being computationally fast with no need for gradient-based optimization techniques. It provides meaningful and informative projections with respect to the response variables, while also preserving a large number of the intrinsic variability of the data.

Nonetheless, there remains potential for further improvement. While we have tested CSPCA's performance in non-linear simulations and complex real datasets, it remains by definition, a linear dimensionality reduction technique. While the inclusion of the delta kernel to model the response variables helps mitigate this limitation for classification tasks, CSPCA is still not guaranteed  to perform equally well in complex non-linear datasets for regression tasks. For this reason, our future work involves a kernelised extension to CSPCA to model non-linear relationships between the data and the response variables in regression tasks. This could also enhance CSPCA's performance in data visualization tasks. Moreover, since CSPCA as presented here is a deterministic method, it could be extended into a probabilistic framework similar to that of probabilistic PCA and probabilistic SPCA \citep{Tipping2002Probabilistic-P, Yu2006Supervised-prob}, through the use of latent variable models. This could allow for the use of Bayesian methods for making inference. Finally, in our extension of CSPCA, we used the standard Nyström's method for approximating the eigenvectors of the matrix of interest. More sophisticated low-rank approximation methods can be considered, apart from variations of Nyström's method, to approximate the eigenvectors and potentially improve the performance of CSPCA even further for large high-dimensional datasets. A key direction for our future research involves integrating sparsity into the CSPCA framework to derive sparse projections, enabling simultaneous dimensionality reduction and variable selection.

\section*{Data Availability Statement}
All data generating mechanisms, simulation analyses and software implementation for the real world analyses are publicly available in the following repository: \textit{\url{https://github.com/theopapazoglou/CSPCA.git}}. All real world datasets used in this paper are publicly available as discussed in the experimental section either through the UCI Machine Learning Repository or through R's Bioconductor. All software implementation was performed using Python.

%Bibliography
\bibliographystyle{plainnat}  
\bibliography{CSPCA_Arxiv}

\end{document}